\documentclass[pra,twocolumn,tightenlines,nofootinbib]{revtex4-1}
\usepackage{bm,dcolumn,amsmath,graphicx,amsfonts,amssymb}

\usepackage{hyperref} 
\usepackage{xcolor}
\definecolor{cite}{rgb}{0.,0.,0.85}   
\hypersetup{colorlinks,linkcolor={cite},citecolor={cite},urlcolor={cite}}

\newcommand{\abs}[1]{\ensuremath{\left |#1\right |}}
\newcommand{\braket}[1]{\ensuremath{\langle #1\rangle}}	

\renewcommand{\v}[1]{\ensuremath{\boldsymbol{#1}}}		
\newcommand{\vv}[1]{\ensuremath{\boldsymbol{\vec{#1}}}}		

\DeclareMathOperator\erf{erf}

\def\d{\ensuremath{{\rm d}}}
\def\deg{{^\circ}}
\def\h{\ensuremath{\hbar}}

\renewcommand{\a}{\ensuremath{\alpha}}

\newcommand{\s}{\ensuremath{\sigma}} 
\renewcommand{\t}{\ensuremath{\tau}} 
\renewcommand{\k}{\ensuremath{\kappa}}

\def\CT{\ensuremath{{\cal T}}}
\def\L{\ensuremath{{\cal L}}}

\newcommand{\un}[1]{\ensuremath{\,{\rm{#1}}}} 

\renewcommand{\bar}[1]{\overline{#1}}
\newcommand{\Geff}{\ensuremath{\Gamma_{\rm eff}}}
\newcommand{\Nclk}{\ensuremath{N_{\rm clk}}}

\begin{document}

\title{Search for transient ultralight dark matter signatures with networks of precision measurement devices using a Bayesian statistics method}

\author{B. M. Roberts}
\author{G. Blewitt}
	\affiliation{Department of Physics, University of Nevada, Reno, 89557, USA}
\author{C. Dailey}
	\affiliation{Department of Physics, University of Nevada, Reno, 89557, USA} 
\author{A. Derevianko}
	\affiliation{Department of Physics, University of Nevada, Reno, 89557, USA}

\date{ \today }  

\begin{abstract}
We analyze the prospects of employing a distributed global network of precision measurement devices as a dark matter and exotic physics observatory. 
In particular, we consider the atomic clocks of the Global Positioning System (GPS), consisting of a constellation of 32 medium-Earth orbit satellites equipped with either Cs or Rb microwave clocks and a number of Earth-based receiver stations, some of which employ highly-stable H-maser atomic clocks.
High-accuracy timing data is available for almost two decades.
By analyzing the satellite and terrestrial atomic clock data, it is possible to search for transient signatures of exotic physics, such as ``clumpy'' dark matter and dark energy, effectively transforming the GPS constellation into a 50,000\,km aperture sensor array.
Here we characterize the noise of the GPS satellite atomic clocks, describe the search method based on Bayesian statistics, and test the method using simulated clock data.
We present the projected discovery reach using our method, and demonstrate that it can surpass the existing constrains by several order of magnitude for certain models.
Our method is not limited in scope to GPS or atomic clock networks, and can also be applied to other networks of precision measurement devices.
\end{abstract}

\setcounter{tocdepth}{1}

\maketitle

\section{Introduction}\label{sec:intro}

Astrophysical observations suggest that ordinary luminous and baryonic matter contributes only about 5\% to the total energy density of the Universe, with the rest due to dark matter (DM)  at $\sim 25\%$, and dark energy (DE) at $\sim 70\%$.
Despite the overwhelming cosmological evidence for the existence of DM, and the considerable effort of the scientific community over several decades, there is as of yet no definitive evidence for DM in terrestrial experiments.

Currently all the evidence for DM comes from observations carried out over distances greater than or comparable to  galactic  scales \cite{Bertone2005}.  
In general, in order to perform  a direct DM detection experiment, these vast 10\,kpc ($\sim10^{13}\,$m)  distances must be extrapolated down to scales that are accessible in laboratory settings ($\sim$\,1\,m). 
This extrapolation leads to a  variety  of plausible theoretical possibilities for DM models, ranging from elementary particles to black holes. 
Considering the broad variety of models and the associated assortment of non-gravitational interactions of DM with ordinary matter, it is important to constrain DM models by creatively reanalyzing archival data~\cite{BudkerPT2015}.
Compared to investments into dedicated experiments, this is a relatively low-cost strategy with potential for important discovery.
Here we develop a method based on Bayesian statistics for a time-domain DM search using data accumulated by networks of precision measurements devices.

The field of low-energy precision measurements has proven to be an important area for probing fundamental laws and searching for new physics, that is often complementary to collider experiments~\cite{AtomicReview2017}.
The idea of using a distributed network of precision measurement devices to search for DM and other exotic physics signatures is one promising approach~\cite{Pospelov2013,Pustelny2013,DereviankoDM2014,GPSDM2017,Delva2017}.
The particular network considered here is the Global Positioning System (GPS), a satellite constellation of nominally 32 satellites in medium-Earth orbit (altitude $\sim20,000$\,km) housing atomic clocks, as well as a large number of atomic clocks on ground-based receiver stations. 
Following the proposal of Ref.~\cite{DereviankoDM2014}, we use the GPS constellation as a $\sim50,000\,$km aperture sensor array, analyzing the satellite and terrestrial atomic clock data for transient signatures of exotic physics, such as DM and DE.
High-quality timing data from the GPS network exists for the past 18 years, and is made freely available by, e.g., the Jet Propulsion Laboratory (JPL), NASA \cite{[{Data freely available online:~}][]JPLigsac,MurphyJPL2015}. This dataset is routinely augmented with more recent data.

The global scale of the GPS network offers a unique opportunity to search for spatially-extended DM objects (or ``clumps''), such as topological defects (TDs)~\cite{Vilenkin1985}, which are otherwise not detectable by most  ongoing and planned DM searches.
The large number of clocks and the very large aperture of the network increase both the chance of an interaction and the sensitivity of the search, since we seek the correlated propagation of new physics signals throughout the entire network.
The large network diameter also increases the overall interaction time.
Therefore, by analyzing the  GPS timing data, one can perform a sensitive search for transient signals of exotic physics, and if no sought signals are found, stringent limits on the relevant interaction strengths can be placed.

Recently, our GPS.DM collaboration carried out an initial analysis~\cite{GPSDM2017} of the archival GPS data, looking for signatures of a particular type of TDs (domain walls, quasi-2D cosmic structures). 
While no such signatures were found, we placed limits on certain DM couplings to atoms that are many orders of magnitude more stringent than the previous constraints.
Here, we present a search method based on Bayesian statistics. We demonstrate that compared to our initial search, the Bayesian approach greatly increases the search sensitivity. This approach also broadens the discovery reach to more general DM models and to lower DM field masses.

Our approach is not limited in scope to the GPS network, but applies equally to other networks of precision measurement devices.
In principle, timing data from any other atomic clocks as well as data from other precision measurement devices can be included in the analysis.
In particular, there are similarities to another experiment, the Global Network of Optical Magnetometers for Exotic physics (GNOME), which employs a geographically-distributed Earth-based network of magnetometers to search for transient signatures of exotic physics, including topological defect DM \cite{Pospelov2013,Pustelny2013}.
Techniques described in this work may prove useful for such experiments.

Beyond ``clumpy'' DM models, one can use  networks to search for other types of DM, such as non-self-interacting virialized ultralight fields (VULFs), that lead to signals that oscillate at the DM Compton frequency.
Such a search would rely on a multi-node spatio-temporal correlation function \cite{DereviankoVULF2016}.
One may also search for both transient and oscillating effects due to ultralight DM (and other exotic physics) with laser interferometers and gravitational wave detectors
\cite{
Arvanitaki2014,Arvanitaki2015,StadnikLaser2015,*StadnikLasInf2015,StadnikDMalpha2015,Arvanitaki2016,Hall2016,Yang2016}, 
by directly exploiting the scalar--photon coupling
\cite{Sikivie1983,
ADMX2010,
CAST2015,*CAST2017
},
 atomic spectroscopy \cite{Tilburg2015,Hees2016,Kalaydzhyan2017} and
noise statistics \cite{RobertsAsymm2018,Kalaydzhyan2018},
 electric dipole moment searches
\cite{Graham2013,Budker2014,RobertsCosPRD2014,*RobertsCosmic2014},
and even pulsar timing \cite{StadnikDefects2014}.

The structure of this paper is as follows.
Section~\ref{sec:theory} reviews the background theory for topological defect DM, the DM-induced transient variations of fundamental constants, and how atomic clocks can be used to search for DM signatures.
Section \ref{sec:noise} discusses aspects of the GPS network relevant to our search.
In Section \ref{sec:SearchMethod} we describe the Bayesian statistics method for the data analysis and the search, and in Section \ref{sec:TestMethod} we use this method with simulated data to demonstrate its efficacy.
Finally, in Section \ref{sec:results} we present the projected sensitivity and the discovery reach of the search.

This paper has three appendices, which include the derivation of the  velocity distributions for macroscopic DM objects, a brief characterization of the noise properties of the clock data relevant to our search, and the expected signals for a few specific DM models.
The supplementary information~\cite{[{See the Supplementary Information ``GPS satellite clock noise characteristics'' in the {\em ancillary files} section of this paper's arXiv page}][]Supplement}
 presents a detailed analysis of noise characteristics such as Allan variance, power spectrum, and autocorrelation  for individual GPS satellite clocks. 
Since the intended audience includes both atomic and particle physics communities, we restore $\hbar$ and $c$ in the formulas. We use the rationalized Heaviside-Lorentz units for electromagnetism.

\section{Theory}\label{sec:theory}

\subsection{Ultralight dark matter and topological defects}\label{sec:TDs}

Despite the extensive searches, both laboratory direct detection and high-energy collider experiments have so far failed to yield convincing evidence for the existence of weakly interacting massive particles (WIMPs) with masses $\sim$\,10\,--\,$10^4\un{GeV}$, see, e.g., 
Refs.~\cite{AgneseSCDMS2014,
PandaX2016,
LUXjan2017,
XENON100-2016,XENON-1T-2017}.
While WIMPs are theoretically well-motivated, they are by no means the only DM candidate.
The null WIMP searches have partially motivated searches for ultralight bosonic DM, such as axions \cite{Peccei1977a,*Peccei1977b,Srednicki1981,Sikivie1983,Preskill1983}.
While direct DM searches with particle detectors rely on measuring energy deposition by individual DM particles, precision measurement techniques are well suited for detecting candidates that act as coherent entities on the scale of individual devices or their networks. 
In other words, precision measurement devices can be used for detecting ultralight DM and this approach probes the mass region that is
complementary to particle detectors.

Ultralight fields may form coherent (on certain time-scales) oscillating fields, or they may form stable macroscopic-scale objects 
\cite{
Sikivie1982,
Press1989,
Vilenkin1994,
Battye1999,
Durrer2002,
Friedland2003,
Avelino2008%
}. The formation of macroscopic ``clumpy'' DM objects requires self-interactions in the dark sector.
An example of macroscopic DM are topological defects, which may have various dimensionalities: monopoles (0D), strings (1D), and domain walls (2D).  Depending on their cosmological fluid equation of state, these objects can contribute to both DM and DE.

The interactions of light scalar fields with standard model (SM) fields can be phenomenologically parameterized as a sum of effective interaction Lagrangians (portals) \cite{DereviankoDM2014}
\begin{equation}\label{eq:effL}
\L_{\rm int} = \L^{\rm PS} + \L^{\rm S^1} + \L^{\rm S^2}  + \; \ldots \; ,
\end{equation}
where $\L^{\rm PS}$ represents the pseudoscalar (axionic) portal, and $\L^{\rm S^1}$ and $\L^{\rm S^2}$ are the linear and quadratic scalar portals, respectively.
The linear and quadratic scalar portals can lead to changes in the effective values of certain fundamental constants and thus cause shifts in atomic transition frequencies.  Atomic clocks in particular are sensitive probes of varying fundamental constants.
The axionic portal leads to interactions that mimic spin-dependent shifts due to fictitious magnetic fields, and  thus are well suited for magnetometry searches \cite{Pospelov2013,*Pustelny2013,KimballQball2017}.
We also note that there are stringent limits on the interaction strength for the linear scalar interaction coming from astrophysics and gravitational experiments (see, e.g., \cite{Raffelt1999,Bertotti2003}).
However, the constraints on the quadratic portal are substantially weaker \cite{Olive2008}.
For concreteness, here we will focus on the quadratic scalar portal.

While we  refer to specific models, namely topological defect DM with quadratic scalar couplings, it is important to note that the search technique is not limited in scope to this possibility.
Any large (on laboratory scales), ``clumpy'' object that interacts with standard model particles is detectable using this scheme.
Examples of such other models include $Q$-balls~\cite{Coleman1985,Kusenko2001,Lee1989}, solitons~\cite{Marsh2015,Schive2014}, axion stars~\cite{Hogan1988,Kolb1993}, and other stable objects formed due to self-interactions in the DM sector.

\subsection{Searching for dark matter with atomic clocks}

Since the microscopic nature of DM is unknown, we take a phenomenological approach for the non-gravitational interactions with ordinary matter~(see, e.g., \cite{DereviankoDM2014}).
Explicitly for the quadratic scalar portal, we have 
\begin{equation}
\label{eq:scalarPortal}
-\L^{\rm S^2}=  \phi^2\left(
{\Gamma_f}{m_f c^2 \bar \psi_f \psi_f   } 
 +\Gamma_\alpha\frac{F_{\mu\nu}^2 }{4}
+ \,\ldots \right),
\end{equation}
where $\phi$ is the scalar DM field (measured in units of energy), $m_f$ are the fermion masses,
$\psi_f$ and $F_{\mu\nu}$
are the SM fermion fields and the electromagnetic Faraday tensor, respectively, 
and  $\Gamma$  are coupling constants that quantify the strength of the DM--SM interaction. 
There is an implicit sum over the SM fermions $f$ in the above equation.
The above Lagrangian leads to the effective redefinition of  fundamental masses and coupling constants,
\begin{align}
\label{eq:varalpha}
\alpha^{\rm eff}(\v{r},t) &= \left[1+ \Gamma_{\alpha}\,{\phi^2(\v{r},t)}\right]  {\alpha },\\
m_{f}^{\rm eff}(\v{r},t) &= \left[1+ \Gamma_{f}\,{\phi^2(\v{r},t)}\right] {m_{f}},
\label{eq:varqcd}
\end{align}
where $\alpha\approx1/137$ is the electromagnetic fine-structure constant 
and $m_f$ are the fermion (electron $m_e$ and light quark $m_q\equiv[m_u+m_d]/2$) masses.
The coupling constants $\Gamma$ have units of  $[{\rm Energy}]^{-2}$ and to aid the comparison with previous literature we also define the effective energy scales  $\Lambda_X \equiv 1/\sqrt{\abs{\Gamma_X}}$  with $X=\alpha,\,m_e,\,m_q$.

Considering TDs, the DM field $\phi^2\to0$ outside the defect, hence the effective couplings are only realized inside the defect.\footnote{Strictly speaking, this condition requires an auxiliary DM field, see Ref.~\cite{DereviankoDM2014} for the mechanism. }
The  field amplitude inside the defect, $A$, can be linked to the average energy density inside the defect as $\rho_{\rm inside}= A^2/(\h c \, d^2)$, where $d$ is the spatial extent or width of the defect. 
In TD models, the width $d$ is set naturally by the field Compton wavelength, $d=\hbar/(m_\phi c)$, where $m_\phi$ is the mass of the DM field particles; in general, we treat $d$ as a free observational parameter.
Further, in the assumption that these objects saturate the local DM energy density, one can link $A$ and $d$ to the local DM energy density $\rho_{\rm DM}$,
\begin{equation}\label{eq:A2}
A^2 = (\hbar c) \, \rho_{\rm DM} v_g  \CT d  ,
\end{equation}
where 
$\CT$
is the average time between close encounters of the DM objects with a point-like device, and the galactic velocity $v_g=\braket{v}\sim300\un{km}\un{s}^{-1} \sim 10^{-3} c$ is the average relative velocity of DM objects that cross paths with the Earth.

From Eqs.~(\ref{eq:varalpha}) -- (\ref{eq:varqcd}), we may relate the observable DM-induced atomic frequency shift to the transient variation of fundamental constants (and thus to the DM field parameters).
The fractional shift in the  frequency $\omega_0$ of a particular clock transition can be expressed as
\begin{equation}
\label{eq:variation}
 \frac{\delta \omega(\v{r}, t)}{ \omega_0} 
= \sum_X \k_X\Gamma_X{ \phi^2(\v{r}, t)}{} \equiv \Geff\,\phi^2(\v{r}, t), 
\end{equation}
where $X$ runs over relevant fundamental constants, and $\k_X$ are dimensionless sensitivity coefficients.
For convenience, we introduced the effective constant, $\Gamma_{\rm eff} \equiv \sum_X \kappa_X \Gamma_X$, which depends on the specific clock.

The dimensionless sensitivity coefficients $\k_X$ are known from atomic and nuclear structure calculations.
For example, considering only the variation in the fine-structure constant $\alpha$ and ignoring relativistic effects, the optical and microwave transitions frequencies scale as
$\omega^{\rm opt}_c \propto \alpha^2$, and
$\omega^{\rm mw}_c \propto \alpha^4$, respectively.
Relativistic atomic-structure effects add small corrections to these scalings~\cite{Dzuba2003,*Angstmann2004}.
For the microwave  Rb, Cs, and H clocks of the GPS network, the effective coupling constants read (using computations~\cite{Dzuba2003,*Angstmann2004,Flambaum2006a,*Dinh2009})
\begin{align}
\label{eq:Crb}
\Geff({\rm ^{87}Rb})&={4.34}\,{\Gamma_\a}-{0.069}\,{\Gamma_{m_q}}+{2}\,{\Gamma_{m_e}},
\\
\label{eq:Ccs}
\Geff({\rm ^{133}Cs})&={4.83}\,{\Gamma_\a}-{0.048}\,{\Gamma_{m_q}}+{2}\,{\Gamma_{m_e}},
\\
\label{eq:Ch}
\Geff({\rm ^{1}H})&={4}\,{\Gamma_\alpha}-{0.150}\,{\Gamma_{m_q}}+{2}\,{\Gamma_{m_e}}.
\end{align}
The values of $\k_{m_q}$ come from a combination of shifts in the nuclear magnetic moment and  in the nuclear size~\cite{Flambaum2006,
[{For Cs, the contributions from these two effects are roughly equal in magnitude and opposite in sign ($\k_\mu+\k_{\rm hq}=0.009-0.007$ \cite{Dinh2009}), and this part of $\k_q$ for Cs is therefore sensitive to uncertainties in the nuclear structure calculations}][]bibtexnote}, and from the variation in the proton mass with $\delta m_p/m_p = 0.05\,\delta m_q/m_q$ \cite{Flambaum2004}.

Although each clock type is sensitive to a combination of three coupling constants,
by combining results for three types of clocks within the network one can unfold individual coupling constants $\Gamma_X$ or, equivalently, individual energy scales $\Lambda_X$. 
Until recently, the existing constraints came from observations of supernova emission~\cite{Olive2008}:
$\Lambda_{m_e,\a}\gtrsim3\un{TeV}$, and 
$\Lambda_{m_p}\gtrsim15\un{TeV}$.
More stringent constraints for certain regions of the $(d,\CT)$ parameter space have recently been placed on $\Lambda_{\a}$ using a laboratory optical Sr clock by the Toru\'n group~\cite{Wcislo2016}. 
Using 16 years of archival GPS data, our GPS.DM collaboration constrained $\Lambda_{\a}, \Lambda_{m _e}$, and $\Lambda_{m_q}$~\cite{GPSDM2017}; 
that initial search focused on domain walls. 
These newly-established constraints reach the $\sim 10^7\un{TeV}$ level depending on the size of the objects and the frequency of their encounters with the Earth.

With the model-specific theoretical background established, now we proceed to developing a method for a sensitive search for  macroscopic DM objects. 
We will demonstrate that  compared to our initial search~\cite{GPSDM2017}, the developed method improves the sensitivity by several orders of magnitude, and also substantially increases the range of probed DM field masses. 
It is also sufficiently general to enable mining for signatures of all the prototypical topological defects: monopoles, strings, and walls.
The method is Bayesian in nature and we start with describing known DM halo properties, velocity distribution and directionality, that serve as priors to the search.

\subsection{Priors on velocity distribution and event rate}
\label{Sec:Priors}

\begin{figure} 
\includegraphics[width=0.3\textwidth]{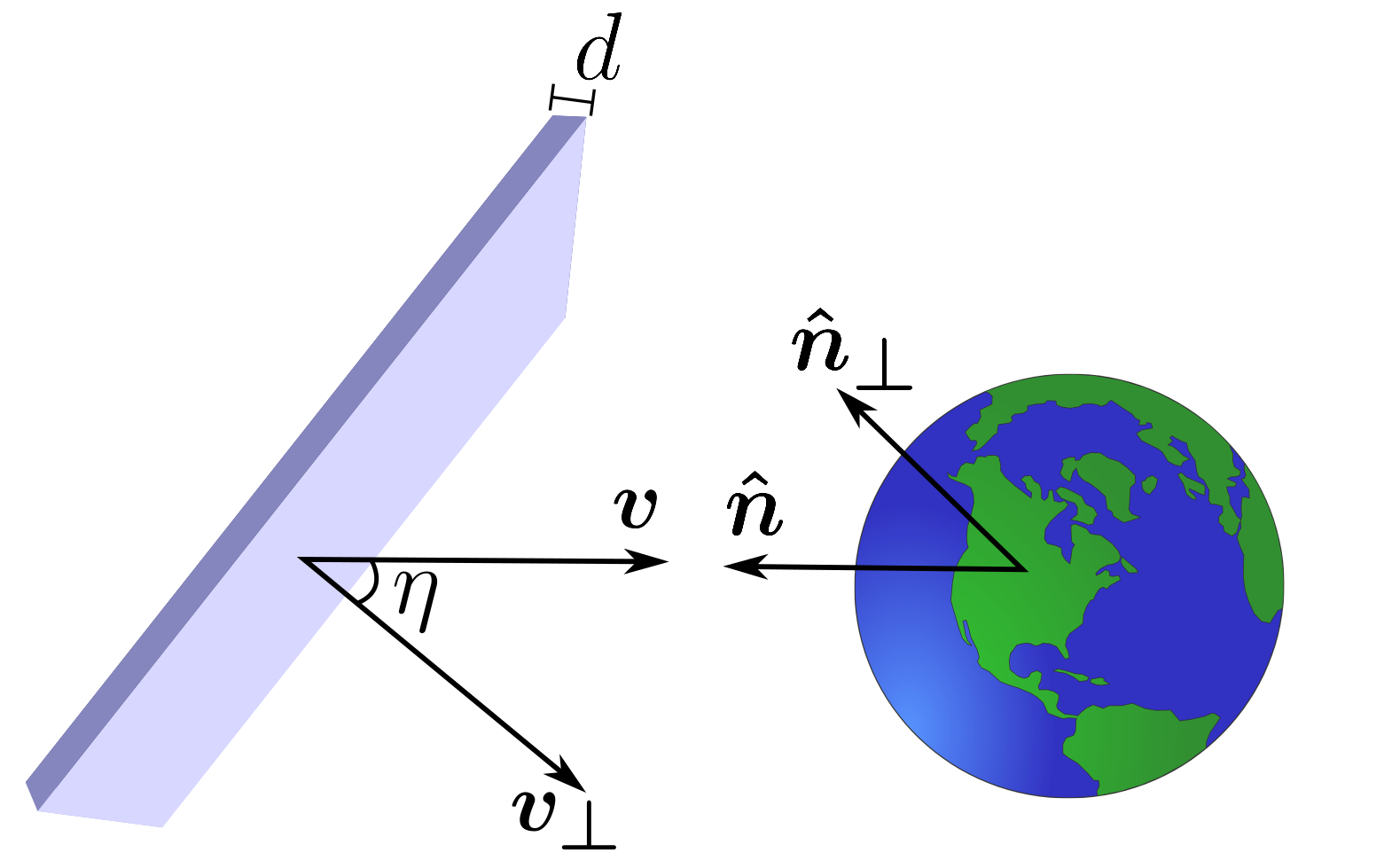} 
\caption{\small  Geometry of a domain wall encounter with the Earth.  
Here, $d$ is the domain wall width, $\v{v}$ is the  relative velocity of the encounter with component $\v{v}_\perp$ perpendicular to the wall surface, and $\eta$ is the angle between $\v{v}$ and $\v{v}_\perp$.  
The incident direction of the wall, $\v{\hat n}$, is defined as pointing away from the Earth center, so that $\v{\hat n} = -\v{v}/\abs{\v{v}}$ and $\v{\hat n}_\perp \equiv -\v{v}_\perp/\abs{\v{v}_\perp}$.
}\label{fig:vperp}
\end{figure}

\begin{figure*}
\includegraphics[width=0.30\textwidth]{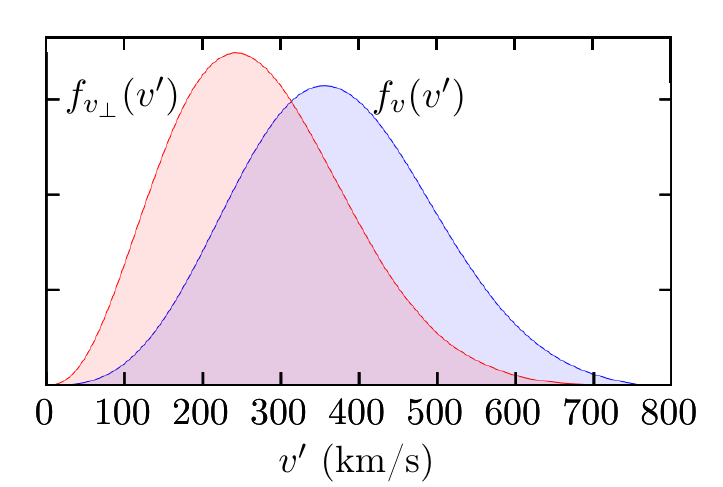}~~~
\includegraphics[width=0.30\textwidth]{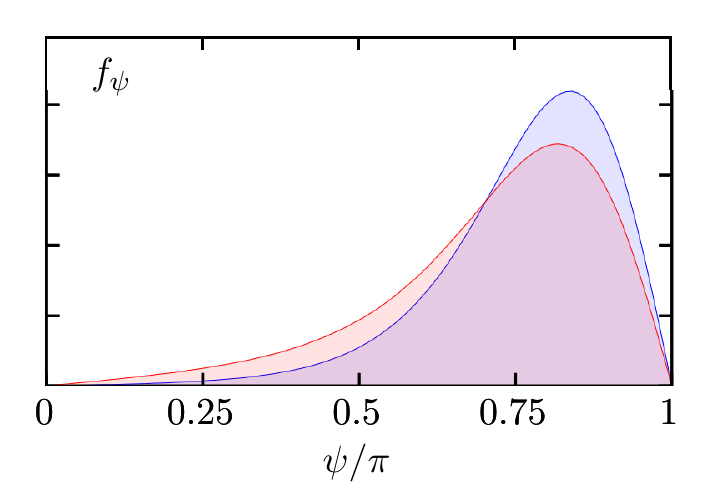}~~~
\includegraphics[width=0.30\textwidth]{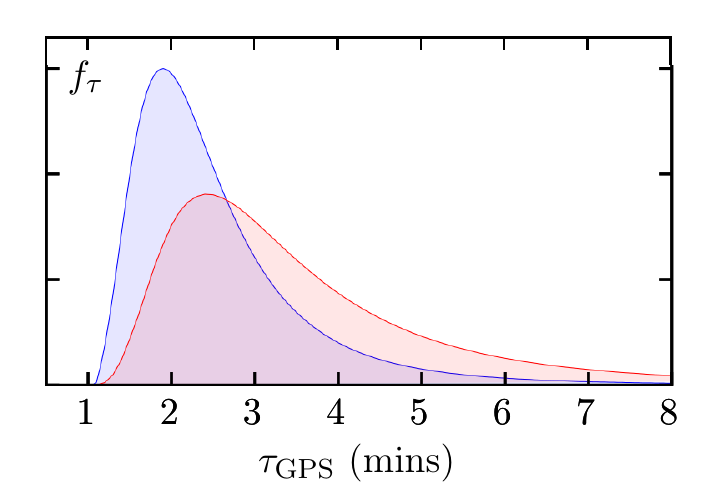}
\caption{\small Probability densities for the DM scalar velocity (left), incident angles (middle), and transit time for the GPS constellation (right).
The forward-facing angle $\psi=\pi$ points in the direction of the galactic motion of the Solar system, towards the Cygnus constellation.
The blue curves show the  standard halo model distributions (relevant for monopole-like DM objects), and the red curves show the distributions for velocities perpendicular to the wall (relevant for domain walls and strings).
}
\label{fig:distros}
\end{figure*}

We form our search priors based on the standard halo model (SHM), see, e.g., Ref.~\cite{Bovy:2012tw}.
Within the SHM framework, the velocity distribution of DM objects in the galactic rest frame is isotropic and quasi-Maxwellian; further details are given in Appendix~\ref{sec:vel-distro}.
The Milky Way rotates through the DM halo, with the Sun moving at $v\simeq 220\un{km}\un{s}^{-1}$ in the direction towards the Cygnus constellation.
This defines the most probable incident direction for a collision with a DM object; in fact, more than 90\% of events are expected to come from the forward-facing hemisphere, as shown below. 
We define the unit-vector, $\v{\hat n}_g$, that points from the Earth center along this direction.
Further, as shown in Fig.~\ref{fig:vperp}, we define the incident direction of the DM object, $\v{\hat n}$, to be pointing away from the Earth center, so that $\v{\hat n} = -\v{v}/\abs{\v{v}}$, where $\v{v}$ is the velocity vector of the DM object.
The  angle of incidence $\psi$ is defined as $\v{\hat n} \cdot \v{\hat n}_g=-\cos\psi$. According to this definition, the forward-facing angle $\psi=\pi$ points in the direction of the galactic motion of the Solar system, towards the Cygnus constellation.

We consider three topological defect ``templates'': domain walls, strings, and monopoles.
We assume that over the length scales of the GPS constellation a string/wall can be modeled to be straight/flat.
A domain wall that crosses the GPS constellation incident with a velocity $\v{v}$ that is at an angle $\eta$ to the vector normal to the wall, would be indistinguishable from a wall (of the same width) incident with a (slower) velocity 
\begin{equation}\label{eq:eta}
v_\perp=v\cos\eta
\end{equation}
that is normal to the wall, see Fig.~\ref{fig:vperp}. 
We will refer to the $\v{v}_\perp$ component of the relative velocity $\v{v}$ as the ``perpendicular'' velocity, and define $\v{\hat n}_\perp \equiv -\v{v}_\perp/\abs{\v{v}_\perp}$.
The same argument applies to strings (for strings, $\v{v}_\perp$ is defined to lie in the plane containing $\v{v}$ and the string symmetry axis).
Therefore, in these cases, the more relevant quantity is the distribution of the perpendicular velocities, $f_{v_\perp}$; this distribution is derived in Appendix~\ref{sec:vel-distro}.

Since we are focusing on macroscopic DM objects, it is also instructive to consider the distribution of transit durations. 
The transit duration, $\t=d/v$, is defined as the time it takes a DM object of width $d$ to sweep through a point in space (or a single device). 
Similarly we can consider $\t_{\rm GPS}=D_{\rm GPS}/v$, the time for the center of the DM object to pass the entire GPS constellation.
Our derived speed, incident angle, and transit duration distributions are shown for monopole- and wall-like objects in Fig.~\ref{fig:distros}.

We treat the expected event rate, $1/\CT$ [see Eq.~(\ref{eq:A2})], as a free parameter.
This parameter can be linked to the number density of DM objects in the galaxy. 
For monopole-like objects (including non-topological solitons, Q-balls, bubbles etc.), the relevant quantity is the volume number density, while for strings and domain walls it is the areal and linear number densities, respectively.
Thereby, $\CT$ can be related to the energy density inside the DM object as
\begin{equation}\label{eq:T}
\CT = \frac{\rho_{\rm inside}}{\rho_{\rm TDM}} \frac{d}{v_g},
\end{equation}
where $\rho_{\rm TDM}$ is the galactic energy density of the considered DM objects.
In the assumption that these objects saturate the local DM density,  we have $\rho_{\rm TDM} = \rho_{\rm DM}$. 
Direct measurements of the local DM density give $0.3\pm0.1\un{GeV}\un{cm}^{-3}$~\cite{Bovy:2012tw}; we take $\rho_{\rm DM}\approx 0.4\un{GeV}\un{cm}^{-3}$ for definitiveness and to be consistent with recent literature.
Note that Eq.~(\ref{eq:T}) is model independent and applies to any DM object of characteristic size $d$.

\section{GPS Architecture and Clock Solutions}\label{sec:noise}

A  detailed description of modern GPS data acquisition and processing techniques and their application in precision geodesy can be found in Ref.~\cite{[][{; see also references therein.}]Blewitt2015307}. Details relevant to DM searches with GPS constellation
are given in Ref.~\cite{GPSDM2017}. Here, we briefly review the main aspects of GPS and introduce  relevant concepts and terminology.

GPS works by broadcasting microwave signals from nominally 32 satellites in medium-Earth orbit (altitude $\sim$\,$20,000$\,km).
The transmissions are driven by an atomic clock (either based on Rb or Cs atoms) on board each satellite. 
It is namely the carrier phase of these microwave signals that is measured by the specialized GPS receivers and is used in deriving the GPS clock solutions.
Typically, each satellite houses four atomic clocks, only one of which is broadcasting at any given time. 
Clock swaps are marked in databases supporting the archival GPS dataset.
There are also a large number of ground-based receiver stations, several of which employ highly-stable H-maser clocks.
The more recent satellites predominantly employ Rb clocks as it has become clear that unpredictable variations in clock phases for the Cs-clock satellites are significantly worse than for Rb. 
As of early September 2017, there were 30 Rb satellites and only two Cs satellites in orbit.

The GPS satellites are grouped into several generations, called blocks: II, IIA, IIR, and IIF~\cite{GPS.gov}, and each satellite is assigned a unique identifier known as the Space Vehicle Number (SVN).
Each subsequent block was built with significant improvements, and the effect of these improvements can be seen in the noise characteristics of the satellite clocks, as discussed below.
Block III satellites are currently under development, and are  to be launched from mid-2018. 
Table~\ref{tab:GPSsats} presents a summary, including the number of days worth of data that is available for each satellite block in the archival data set.
Further, the network can be extended to incorporate the network of Earth-based receiver clocks, as well as clocks from other Global Navigation Satellite Systems, such as the European Galileo, Russian GLONASS, and Chinese BeiDou, and networks of laboratory clocks~\cite{Hees2016,Wcislo2016,Riehle2017,Delva2017}.

\begin{table}%
\centering%
\caption{\small Summary of GPS satellite clocks. The currently employed data set, ranging from 5 May 2000 to 2 September 2017, consists of a total of 186,\,700 clock-days.}%
\begin{ruledtabular}%
  \begin{tabular}{llrrr}%
\multicolumn{1}{c}{Block}&\multicolumn{1}{l}{Years active}&\multicolumn{2}{c}{Days in data}&\multicolumn{1}{l}{In orbit\tablenotemark[1]}    \\
\cline{3-4}
\multicolumn{2}{c}{}&\multicolumn{1}{c}{Cs}&\multicolumn{1}{c}{Rb}&\multicolumn{1}{c}{}    \\
\hline%
I&1978--1995&0 &0 &0\\
II&1989--2007&5304&1181&0\\
IIA&1990--2017&39557&33319&1\\
IIR&1997--&0&92287&19\\
IIF&2010--&2541&12511&12\\
  \end{tabular}%
\end{ruledtabular}%
   \label{tab:GPSsats}%
\tablenotetext[1]{As of September 2017. Also as of that date, only two satellites (both block IIF) use Cs clocks.}
  \end{table}%

Here we analyze data generated by the Jet Propulsion Laboratory (JPL)~\cite{JPLigsac}, in which clock time-series are given at $\tau_0=30\,{\rm s}$ intervals (epochs).
The data are  clock biases, that is the difference in the time readings (clock phases) between the given clock and a reference clock.
The same reference clock is used for the entire GPS network for each day.
The biases are generated using data from a global network of $\sim100$ GPS station receivers \cite{IGSnetwork} by a mature analysis system that is used routinely for purposes of centimeter-level satellite orbit determination, and millimeter-level positioning for scientific purposes, such as plate tectonics, Earth rotation, and geodynamics. 

We also note that while the currently available clock time-series are sampled every 30 seconds, the raw GPS data is sampled every second for some stations. 
It is therefore possible to re-process the GPS data to generate higher-rate 1\un{s} clock solutions. 
This work is currently underway in our group. 
Notice that a fiducial DM object sweep through the entire constellation takes about 170 seconds, thereby it lasts for just 6 epochs for the currently available 30\un{s} sampling intervals.  Clearly, the resolution would improve for the 1\un{s} data.  

In the initial GPS data processing (performed by JPL~\cite{MurphyJPL2015}), there is effectively no restriction on the allowed behavior of the clocks from one epoch to the next.
Crucially, if a clock were to have a real transient that far exceeded engineering expectations, the data over that time window would not have been removed as outliers.

The clock biases from JPL~\cite{JPLigsac} also come with a ``formal error'', $\sigma_F$.
The formal error, typically on the order of $\sigma_F\sim0.02-0.03\,{\rm ns}$, quantifies uncertainty in the determination of the clock bias, and does not directly incorporate the intrinsic clock noise or slowly varying biases due to correlated orbit errors and higher-order general relativistic effects ($\sim0.1\un{ns}$).
Only the most recent satellite clocks (Rb clocks on board the block IIF satellites) have observed temporal variations from one epoch to the next that are at a similar level as the formal error, indicating that temporal variations in older clocks are actually due to clock behavior rather than estimation error.
In fact, the observed variances in the data from the most modern Rb IIF satellite clocks are significantly better than suggested by the formal error, see Appendix~\ref{sec:clocknoise}.

Due to frequency drifts and other long time-scale ($\sim$ hours) effects, it is typical for a second-order polynomial ($y_2$) to be subtracted from the raw GPS time series data before the analysis~\cite{Blewitt2015307}.
One may form $y_2(j)$ for each clock, for each individual day, using a weighted least-squares approach, taking the weights as the inverse of the formal error.
Then the polynomial-reduced data (residuals) are defined
$d_j^{(0)} = x_j-y_2(j)$,
where $\{x_j\}$ are the raw time-series data, and $j$ denotes the same-time (epoch).
This procedure is useful for visualizing the data, however,  it is not necessary for our analysis. 
Unless noted otherwise, we take  $d_j^{(0)} \equiv x_j$ in this paper.

The relative phase of an atomic clock (bias) $d_j^{(0)}$ is a non-stationary time series, dominated by random walk noise.
To perform the analysis, we must first ``whiten'' the data.
To this end, we employ (depending on the clock type, as discussed below) either a first- or second-order differencing, and define
\begin{align}
\label{eq:difference}
d_j^{(1)} &\equiv d_{j}^{(0)}-d_{j-1}^{(0)}, \\
d_j^{(2)} &\equiv d_{j}^{(0)}-2d_{j-1}^{(0)}+d_{j-2}^{(0)}.
\end{align}
In general, first-order differencing is sufficient for Rb clocks, while second-order differencing is required for Cs clocks. 
Since $d_j^{(1)}$ is proportional  to the discreet derivative of the clock biases, we refer to it as a pseudo-frequency.
Further discussion of the clock noise characteristics is presented in Appendix~\ref{sec:clocknoise}; see also the Supplementary Information, where we quantify the noise characteristics of each satellite clock individually.

\section{Bayesian search for DM events}\label{sec:SearchMethod}

\subsection{Likelihoods and odds ratio}\label{sec:likelihoods}

A DM-induced perturbation in the  device data would be indistinguishable from a perturbation caused by other external non-DM factors or random statistical processes.
The key then, is to rely on the correlated propagation of clock ``glitches'' across a network caused by the sweep of a DM object through the network. 
Based on the standard halo model (see Sec.~\ref{Sec:Priors}), DM objects are expected to travel relative to the Earth with galactic-scale speeds, $v_g\sim300\un{km}\un{s}^{-1}$, incident from a certain direction.
Thus the speed and the directionality of the sweeps serve as DM signatures.  
There are similarities between the method we describe and those employed in gravitational wave detection by the LIGO collaboration, see, e.g., Refs.~\cite{Anderson2001,Allen2012}.

Consider a candidate model, denoted $M$, that predicts a DM signal across the network (for example, the passing of a  domain wall). 
In order to determine whether such a model is supported by the data $D$, we employ a Bayesian technique, see, e.g., Ref.~\cite{GregoryBayesian2005}.
In Bayesian statistics, model selection is based on forming the odds ratio of two probabilities (likelihoods)
\begin{equation}
O_{M,\bar{M}}(j_0)=\frac{p(D_{j_0}|M,I)}{p(D_{j_0}|\bar M,I)}.
\label{eq:odds}
\end{equation}
Here, $\bar M$ denotes the proposition that no signal is present in the data, i.e. the data is purely random,
 and $I$ encodes the knowledge of the SHM priors discussed in Sec.~\ref{Sec:Priors}. 
The data stream $D$ is sampled at intervals of $\t_0$ ($\t_0=30\un{s}$ for our current GPS data set).
Since we search for transient events of finite duration, the odds ratio is tested in a time window of length $J_W$ points, centered at epoch $j_0$. 
The value of $J_W$ is determined by the maximum duration of the transient signals to be tested.
In the analysis, we scan over $j_0$ for a fixed value of $J_W$, so the odds ratio is an explicit function of $j_0$. 

The likelihoods $p(D_{j_0}|M,I)$ entering Eq.~(\ref{eq:odds}) are described by the Gaussian multi-variate distributions marginalized over model parameters,
\begin{widetext}
	\begin{align}
		\label{eq:m-likelihood}
		p(D_{j_0}|M,I)&=
		K\int\d^3v \; p(\v{v}|M,I)
		\int\d h\; p(h|M)\;
               \int\d x\; p(x|M)\;   
		\int_{ (j_{0}-1)\tau_0}^{ j_0\tau_0}\frac{1}{\t_0}\,\d t_0\;
		\exp
		\left(
			-\frac{1}{2}\chi^2(s)
		\right)\,.
	\end{align}
\end{widetext}
Here,
$\v{v}$ is the velocity of the incident DM object in the Earth-centered inertial (ECI) frame,
$h$ quantifies the amplitude of the DM signal,
$x$ stands for the remaining model-specific parameters of the DM object,
and $K$ is a normalization factor.  
Further, $t_0$ is the moment of time at which the DM object passes by the center of the Earth. 
It is assumed to occur in the time interval $\left( (j_0-1)\t_0, j_0 \t_0 \right]$, and we marginalize over $t_0$ in the last integral (\ref{eq:m-likelihood}).
Note that compared to our initial search~\cite{GPSDM2017}, the single-device sweep time ($d/v$) may last longer than $\t_0$.

The data and the model-prescribed DM signal are combined in the argument of the exponential, 
\begin{equation}\label{eq:chis}
\chi^2(s)
	=\sum_{ab}^{\Nclk }\sum_{jl}^{J_W}\left[{d^a_j}-{s^a_j}\right]{({E^{-1}})^{ab}_{jl}}\left[{d^b_l}-{s^b_l}\right], 
\end{equation}
where $E$ is the covariance matrix discussed in the following section. 
Here and below we use the ``upstairs'' indices to label devices, and the ``downstairs'' indices to denote epochs (sampling times).
The indices $a$ and $b$ run over all $\Nclk $ devices in the network, and the indices $j$ and $l$ run over the $J_W$ data points in the  time window.   The device data $d$ notation is generic and it can stand for the singly-- or doubly--differenced clock bias data, Eq.~(\ref{eq:difference}).
Finally, $s^a_j=s^a_j(M,t_0,\v{v},h,x)$ is the model-prescribed DM signal in device $a$ at epoch $j$, discussed in Sec.~\ref{sec:signals}.

Continuing with the discussion of factors entering the likelihood, $p(\v{v}|M,I)$ is the (normalized) probability density for the velocity distribution of DM objects in the ECI frame.
In the case of monopoles, for example, it is reasonable to take this to be given by the SHM.
Likewise, the function $p(h|M)$ is the normalized probability density for the DM signal amplitude in the time series, and is described by a flat prior.\footnote{We note that the normalization for the $h$ prior is arbitrary. For our purposes it is not important, since we do not rely on the actual value of the likelihoods function but rather define some threshold, above which false-positives are sufficiently rare, as discussed in the following sections.}
To calculate the likelihoods, we perform the integral over $h$ analytically (possible because $s$ is linear in $h$, see below), and use a randomized Monte-Carlo integration for the other parameters.

Finally, the likelihood that no signal is present in the data
is given simply by
\begin{equation}
		\label{eq:null-likelihood}
		p(D_{j_0}|\bar{M},I)=K
		\exp
		\left(
			-\frac{1}{2}\chi^2(0)
		\right),
\end{equation}
where the DM signal is set to zero. 
The window size $J_W$ is kept the same as in the  $p(D_{j_0}|M,I)$ computations.

The likelihood functions (\ref{eq:m-likelihood}) and (\ref{eq:null-likelihood}) are calculated for every available epoch $j_0$, and the odds ratios~(\ref{eq:odds}) are formed. Large spikes in the odds ratio as  a function of $j_0$ can indicate potential DM events.

\subsection{Correlations and covariance}
\label{sec:covariance}

The covariance matrix entering Eq.~(\ref{eq:chis}) is defined as 
\begin{equation}\label{eq:Eijkl}
E^{ab}_{jl}\equiv\braket{d^a_jd^b_l},
\end{equation}
where $\braket{\cdots}$ denotes averaging.
To compute its elements, one requires a stationary time series, for which (depending on the clock type) we use either the first- or second-order differenced data (\ref{eq:difference}).
Note that for pure uncorrelated white noise, the covariance matrix is completely diagonal, with elements given by the variances.
In this case, the matrix inversion required for computing 
$\chi^2(s)$ (\ref{eq:chis})  is trivial.
Realistic device noise is, however, correlated. 
Specific to the GPS clocks, additional short-range anti-correlation for individual clocks is introduced by the propagation of the formal error (which is roughly white noise in $d^{(0)}$) through the differencing procedure (\ref{eq:difference}).
Moreover, the underlying clock biases $d^{(0)}$ are the differences between the phases of the given clock and a reference clock. 
Since the reference clock is common to all clocks, biases and the differenced data $d^{(1)}$ and $d^{(2)}$  are correlated between different clocks. 

It is convenient to split the covariance matrix into two contributions, $E=A+B$, where 
\begin{align}\label{eq:E=A+B}
A^{ab}_{jl} &\equiv E^{ab}_{jl}\delta^{ab},\\
B^{ab}_{jl} &\equiv E^{ab}_{jl}(1-\delta^{ab}).
\end{align}
The first term, $A$, represents the correlation between data points for a single clock, i.e.,  auto-correlation.
The $B$ contribution describes the correlations between different clocks, and is referred to herein as the cross-correlation.
The autocorrelation part of the covariance matrix is block diagonal, built from $\Nclk$ independent symmetric $J_W \times J_W$ matrices.
The elements of $A$ depend only on the distance from the diagonal, and can be related to the autocorrelation function $A^a(\Delta t)$ as 
$A^{aa}_{jl} = {(\s^a)}^2 A^a(\Delta t_{jl}),$
where $\Delta t_{jl} = \abs{j-l}\tau_0$ is the lag and $\s^a$ is the standard deviation (see Appendix~\ref{sec:clocknoise}).

Each clock in the network is referenced against a common reference clock.
This adds a common noise component to all the data streams, and is main source of cross-correlations.
Therefore, the time series for each clock can be decomposed as
\begin{equation}\label{eq:dec}
d^a_j = e^a_j + c_j,
\end{equation}
where $c_j$ is the component due to the shared reference clock, and $e^a_j$ is the component unique to clock $a$ ($\braket{e^a_je^b_l}=0$ for $a\neq b$).
Then, it is clear that $B$ depends only on the reference clock, and is independent of $a$, $b$:
\begin{equation}
B^{a,b\neq a}_{jl} = \braket{d^a_jd^b_l} = \braket{c_jc_l} \equiv b_{jl}.
\end{equation}

To calculate the likelihoods, we need to invert the covariance matrix.
First, we note that the Earth-based H-maser clocks used as reference in the JPL data processing are typically much quieter than the  satellite clocks.
 Therefore, the cross-correlation contribution $B$ is typically smaller than $A$, so $B$ can be treated perturbatively. 
Further, we may neglect the even smaller terms $B^{ab}_{jl}$ with $j\neq l$, defining $b_0\equiv B^{ab}_{jj}$.

Thus, we express the inverse of the covariance matrix as
${E}^{-1} = H + W,$
where
\begin{align}
H^{aa}_{jl} &= ({A}^{-1})^{aa}_{jl} ,  \\
W^{ab}_{jj} &\approx \frac{-b_0}{({\s^a\s^b})^2}(1-\delta^{ab}).  \label{eq:EHW} 
\end{align}
Then, Eq.~(\ref{eq:chis}) can be expressed (with $\eta \equiv d-s$) as
\begin{multline}%
\chi^2(s)=
\sum_{a}^{\Nclk }\sum_{jl}^{J_W}\eta^a_j\,{H}_{jl}^{aa}\,\eta^a_l
-
 \sum_{a \neq b}^{\Nclk } \sum_{j}^{J_W} \frac{b_0\, \eta^a_j\,\eta^b_j }{({\s^a\s^b})^2}.
\label{eq:chis2}
\end{multline}

The described approximation holds when the clock noises far exceed that of the reference clock. 
This approximation breaks down if the network includes clocks with noise levels similar to that of  the reference clock.
For example, when including multiple station, Rb-IIF, or laboratory clocks,  Eq.~(\ref{eq:EHW}) is no longer valid.
In this case, we define the weighted mean of all (other) clocks
\[\bar d^{\bar a}_j = \frac{\sum_{b\neq a} d^b_j \, (\s^b)^{-2}}{\sum_{b\neq a} {(\s^b})^{-2}} \approx c_j \pm \sigma/\sqrt{\Nclk} ,\]
which is subtracted from each time series [Eq.~(\ref{eq:dec})]:
\begin{equation}\label{eq:WM}
 d^a_j-\bar d^{\bar a}_j \approx e^a_j \pm \sigma/\sqrt{\Nclk}.
\end{equation}
Here, $\s$ is the typical standard deviation of the clock data.
Each new data stream still contains a common component, $\sim$\,$\sigma/\sqrt{\Nclk}$, however this is small enough so that the above approximation (\ref{eq:EHW}) holds true.
In these cases, the same procedure must be applied also to the expected signals $s^a_j \to s^a_j-\bar s^{\bar a}_j $.

\subsection{Transient dark matter signals}\label{sec:signals}

The likelihood function in Eq.~(\ref{eq:m-likelihood}) requires a model-prescribed DM signal, ${s^a_j}$, for the data streams to be compared against.
The DM signal depends on the assumed coupling strength to the device, and on the kinematics and spatial structure of the DM object (monopole, domain wall, etc.). To quantify the transient signal we need to specify the collision geometry. 
We work in the ECI (Earth-centered inertial) J2000  frame, which has its origin (denoted ECI0) at the center of mass of the Earth, and $z$-axis aligned with Earth's spin axis. 
The $x$-axis is aligned with the mean equinox  at 12:00 Terrestrial Time on 1 January 2000.
The important aspect is that the ECI frame orientation remains fixed in the galactic rest frame, i.e., it does not rotate with the Earth.

Here, we consider three generic and geometrically unique templates: walls, strings, and monopoles. 
Albeit more complex geometries are plausible, such as walls closing on themselves forming cosmic bubbles, the presentation below is sufficient for extending the formalism to such more complex object geometries. 
While the field profile inside the DM object can be arbitrary, we focus on Gaussian profiles.  
Beyond qualitative arguments, the reasons for Gaussian-profiled objects can be also supported by Bayesian logic. 
Indeed, application of the maximum entropy principle to a distribution with the  mean and variance (determined by the defect size $d$ in our case) as the only given information yields the Gaussian distribution~\cite{GregoryBayesian2005}.
In any case, the presented formalism can be applied to arbitrarily-shaped DM object profiles.

We  assume that the linear trajectory and velocity of the DM object are not affected by the gravitational pull of the Earth or the portal couplings to the Earth constituents, and that the shape of the DM object is preserved through the encounter. 
Another assumption is that the DM encounters are well separated, i.e.\ DM objects do not overlap and at most one of them interacts with the entire network at any given time. 
Finally, we assume that objects lacking spherical symmetry do not rotate.

\begin{figure}
\includegraphics[width=0.35\textwidth]{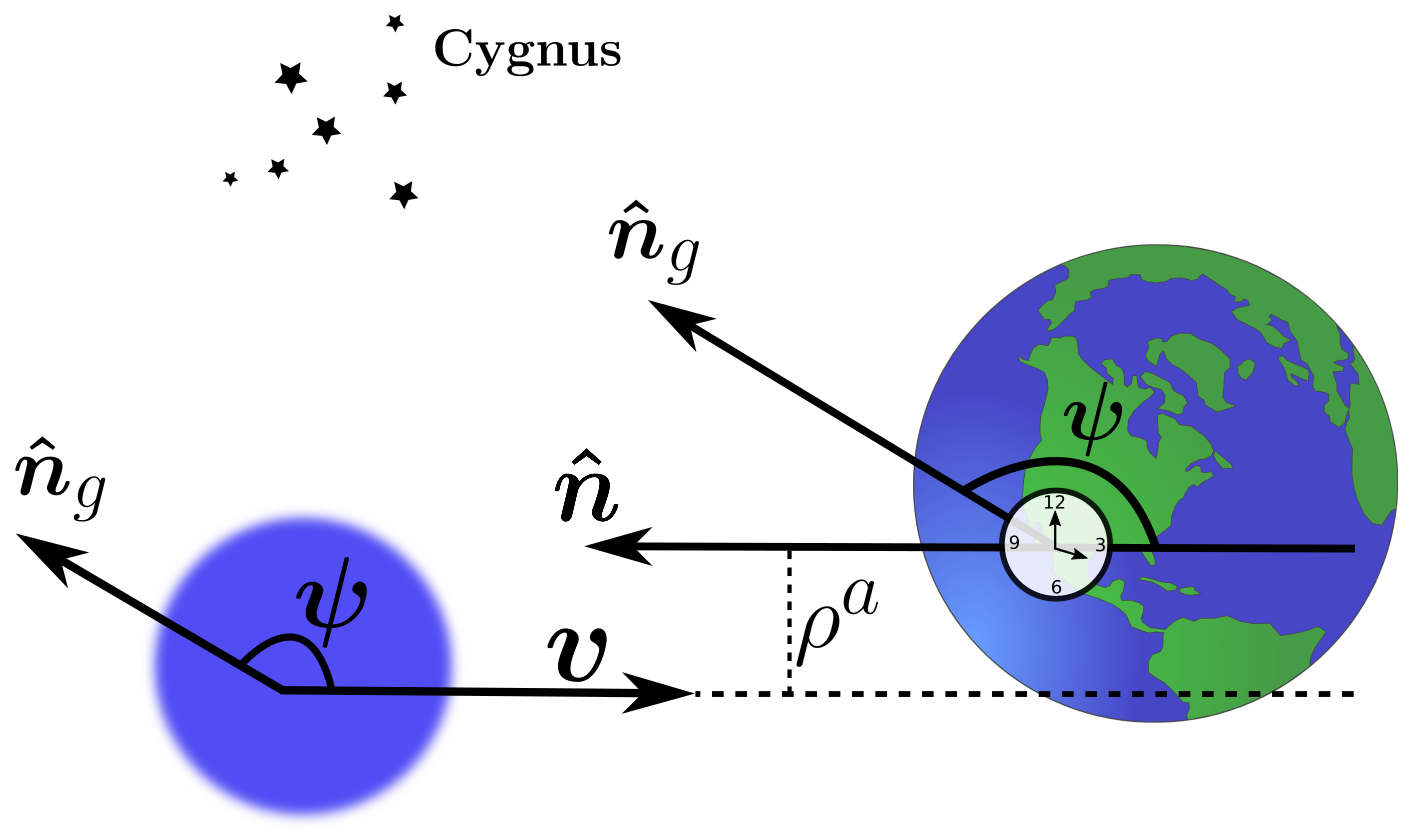} 
\caption{\small Example geometry for a monopole crossing.}
\label{fig:monopole}
\end{figure}

Consider an event in which the center of a DM object moving with velocity \v{v} crosses the plane that is perpendicular to $\v{v}$ and contains ECI0 at time $t_0$, as shown in Fig.~\ref{fig:monopole}.
The accumulated time bias between a clock thats frequency is perturbed by $\delta\omega$ and an unaffected clock ($\omega_0$) is given by $\int_{-\infty}^t\frac{\delta\omega(t')}{\omega_0}\d t'$.
Therefore, at time $t$, the DM-induced clock phase bias in clock $a$ reads
\begin{equation}
\label{eq:generalsignal}
{s^a}^{(0)}(t) =  \int\limits_{-\infty}^t 
		\left[ 
			h^a \varphi_M^2(t^a,{\rho^a},t') - h^R \varphi_M^2(t^R,{\rho^R},t')
		\right] \d t' 	,
\end{equation}
where $\varphi_M^2$ is the normalized profile\footnote{The DM ``profile'' $\varphi$ differs from the field $\phi$ [Eq.~(\ref{eq:scalarPortal})] only by normalization, and is defined for convenience; see Appendix~\ref{sec:specific-signals}.} of the DM object (for specific model $M$), 
${\rho^a}$ is the impact parameter, 
$h \propto A^2\,\Gamma_\mathrm{eff}$ 
is a clock-specific constant that determines the magnitude of the signal in the data (see Appendix~\ref{sec:specific-signals}), and $t^a$ ($t^R$) is the time of encounter for clock $a$ (reference clock).
The time of encounter is defined as the moment the DM object passes by clock $a$.
More precisely, it is the time at which the center of the DM object (central plane for walls, or central axis for strings) crosses the plane that is perpendicular to \v{v} and contains the given clock:
\begin{equation}\label{eq:tiR}
t^a = t_0-\frac{{\v{r}^a}\cdot\v{\hat n}}{v},
\end{equation}
where $\v{\hat n}$ is the unit vector that points from ECI0 parallel to the incident direction of the DM object
($\v{v}=-v\v{\hat n}$, see Fig.~\ref{fig:monopole}), and ${\v{r}^a}$ is coordinate of clock $a$ in the ECI frame. 
The satellite and ground station positions $\v{r}^a$ are a part of the JPL GPS dataset, and are known with  $\sim\mathrm{cm}$ and  $\sim\mathrm{mm}$ accuracies, respectively.
Note that the impact parameters are zero for domain walls, but are, in general, non-zero for strings and monopoles; see Appendix~\ref{sec:specific-signals}.

For domain walls and strings, we use $\v{v}_\perp$ and $\v{\hat n}_\perp$, see Fig.~\ref{fig:vperp} and the discussion around Eq.~(\ref{eq:eta}). 
The discreet matrix ${s^a_j}$ is generated by integrating to the specific values of $t$ that correspond to the GPS epochs (data sampling times).
Then we form either the first- or second-order differenced DM signals as in Eq.~(\ref{eq:difference}), with $d\to s$.

The particular form of $\varphi^2_M$ depends  on the spatial structure and the kinematics  of the DM object.
In Appendix~\ref{sec:specific-signals}, we present explicit signals for domain walls, monopoles, and strings, and link $\varphi_M$ and $h^{a}$ to the field parameters for these templates.

In our discussion of DM signals, we neglected the Earth orbital motion about the Sun at $\sim 30 \un{km/s}$, orbital velocities of satellites about the Earth ($\sim 4 \un{km/s}$), and the ground station rotational velocities ($\sim 0.5 \un{km/s}$). 
While these velocities are much smaller than the galactic velocities, the motional effects can become important for large-scale or slowly-moving objects. 
For example, the motional effects become relevant if the overall duration of an encounter is comparable to the 12-hour satellite orbit.
The modification of the DM signal templates to account for clock motion is straightforward, as the satellite and ground station positions are known. 
We leave this generalization for future work.

\subsection{Mixed networks}

There are several different clock types (Cs, Rb, H-maser) in the GPS network.
As our search is expanded to include other laboratory clocks (and other high-precisions sensors) the diversity will increase further.
Each clock species may respond differently to the interaction with the DM field, see Eq.~(\ref{eq:variation}).
Therefore, we cannot assume $h$ to be uniform  across the network.

There are several approaches for inhomogeneous networks.
One approach, as per Ref.~\cite{GPSDM2017}, is to consider separately the homogeneous sub-networks (e.g., consider only the Rb clocks).
The major drawbacks of this approach is that we lose the benefit of the highly-stable H-maser reference clocks (none of the GPS satellite clocks have H-masers), and that we also limit the total number of clocks that are considered at any given time.

The simplest approach is to assume that one of the couplings in Eq.~(\ref{eq:variation}) dominates, and carry out the analysis separately for each case.
For example, we may assume that $|\Gamma_\alpha|\gg |\Gamma_{m_e}|, |\Gamma_{m_q}|$ in Eq.~(\ref{eq:scalarPortal}).
The drawback of this approach is that it does not account for the possibility that several couplings may produce effects that are of a similar magnitude.

Furthermore, a Bayesian-like approach is to introduce additional marginalization parameters for each extra free parameter in place of $h$.
The number of such free parameters is equal to the smaller of either the number of distinct clock species in the network, or the number of distinct couplings we consider.
For example, considering a network of Rb, Cs, and H clocks (as per GPS), we can substitute 
$\int\d h\to \int\d h_{\rm Rb}\int\d h_{\rm Cs}\int\d h_{\rm H}$ in Eq.~(\ref{eq:m-likelihood}).

\subsection{Directional signatures}\label{sec:directional}

A possible scenario is that  a large number of small events are flagged by the Bayesian search (by ``small'' we mean the magnitude of the signal in the data compared to the clock noise, or the small magnitude of the spikes in the odds ratio).
Of course, such events may be  simply due to random statistical fluctuations, or other conventional-physics non-DM perturbations.
Here we consider signatures unique to DM (or other galactic sources) allowing us to exclude non-DM signals. 
While these signatures are  included in the Bayesian approach through the priors (e.g., the likelihood are suppressed for velocities outside the SHM range through the prescribed velocity distribution prior), we could also examine inferred values of collision geometry parameters through the Bayesian parameter estimation, as discussed in Sec.~\ref{sec:parameters}. 
Being able to resolve the event velocity magnitude and directionality is a powerful feature of geographically distributed networks.

First of all, the distributed nature of the network offers the direct sensitivity to the magnitude of DM object velocities. 
If the observed incident velocity falls too far outside of the bounds allowed by the standard halo model, then a DM origin can be excluded. 
There is also sensitivity to the directionality of the DM object velocity. 
The most probable incident direction is from the average forward direction of the Sun's motion through the galaxy (roughly from the direction of the Cygnus constellation), see Fig.~\ref{fig:distros}.
We are only aware of one external systematic effect that has propagation speeds comparable to $v_g$, which is the solar wind~\cite{SolarWindBook}.
This effect, however, can be vetoed out on the basis of distinct directionality from the Sun, and by the fact that the solar wind does not affect the satellites in the Earth's shadow.

\begin{figure}
	\includegraphics[width=0.40\textwidth]{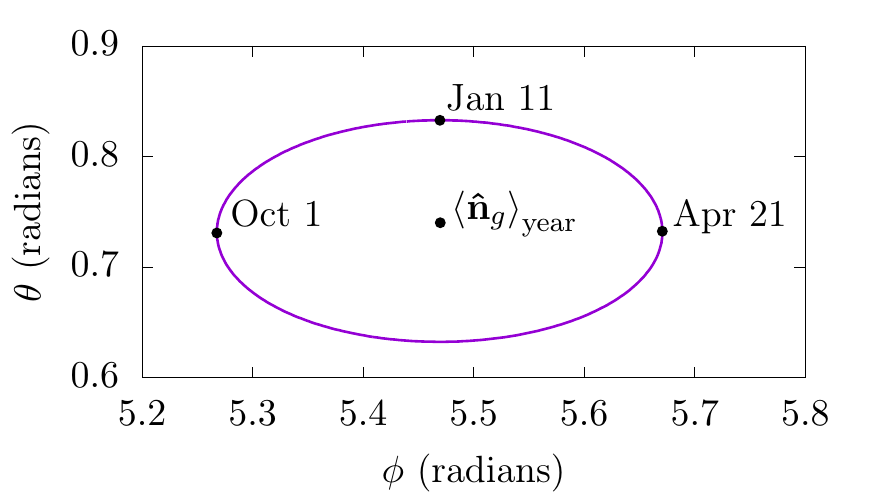}
	\caption{\small Annual variation in the direction of the Earth's galactic motion (ECI frame, $\theta\in[0,\pi]$ is the polar angle), which is the most probable incident DM direction. The central point, $\v{\hat n}_g$, is the average direction, corresponding to the direction of the Sun's velocity through the galaxy.}
	\label{fig:AngleVariation}
\end{figure}

In addition to individual event signatures, one can also focus on the overall event statistics, provided the event rates are sufficiently high on the yearly basis \cite{RobertsAsymm2018}. 
For example, due to the $\sim10\%$ annual variation in the relative velocities of the Earth and Sun in the galaxy~\cite{Freese2013}, one would expect to observe the annual modulation in the event rate.
This approach parallels the method employed in WIMP searches, e.g.,~Refs.~\cite{CoGeNT2011,Bernabei2013}. 
Unlike WIMP searches, where the event rate may depend strongly on the DM velocity~\cite{RobertsDAMA2016,*RobertsAdiabatic2016} (due to energy dependence of the cross section), here the rate is linear in $v$.
Also unlike (most) WIMP searches, the distributed network approach is additionally sensitive to the annual modulation in the average incident velocity {\em direction}, which varies by $\sim$\,$20\deg$, as shown in Fig.~\ref{fig:AngleVariation}.
(A WIMP-detection scheme that does have directional sensitivity is presented in  Ref.~\cite{Rajendran2017}.)

\section{Benchmarking the method}\label{sec:TestMethod}

\subsection{Simulating realistic clock time series}
\label{sec:TestMethod-randts}

\begin{figure*}
	\includegraphics[height=0.26\textwidth]{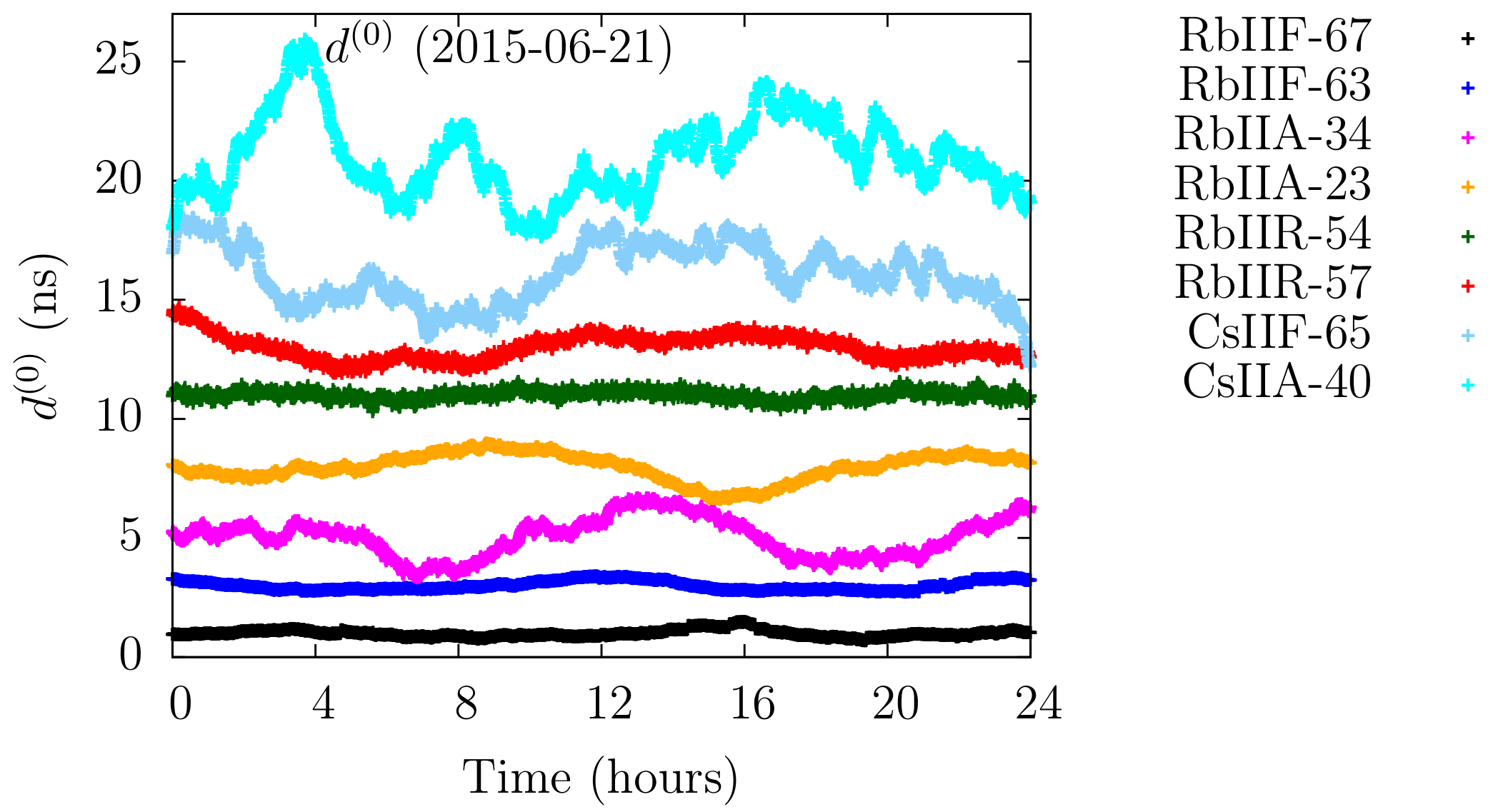}~~~~
	\includegraphics[height=0.26\textwidth]{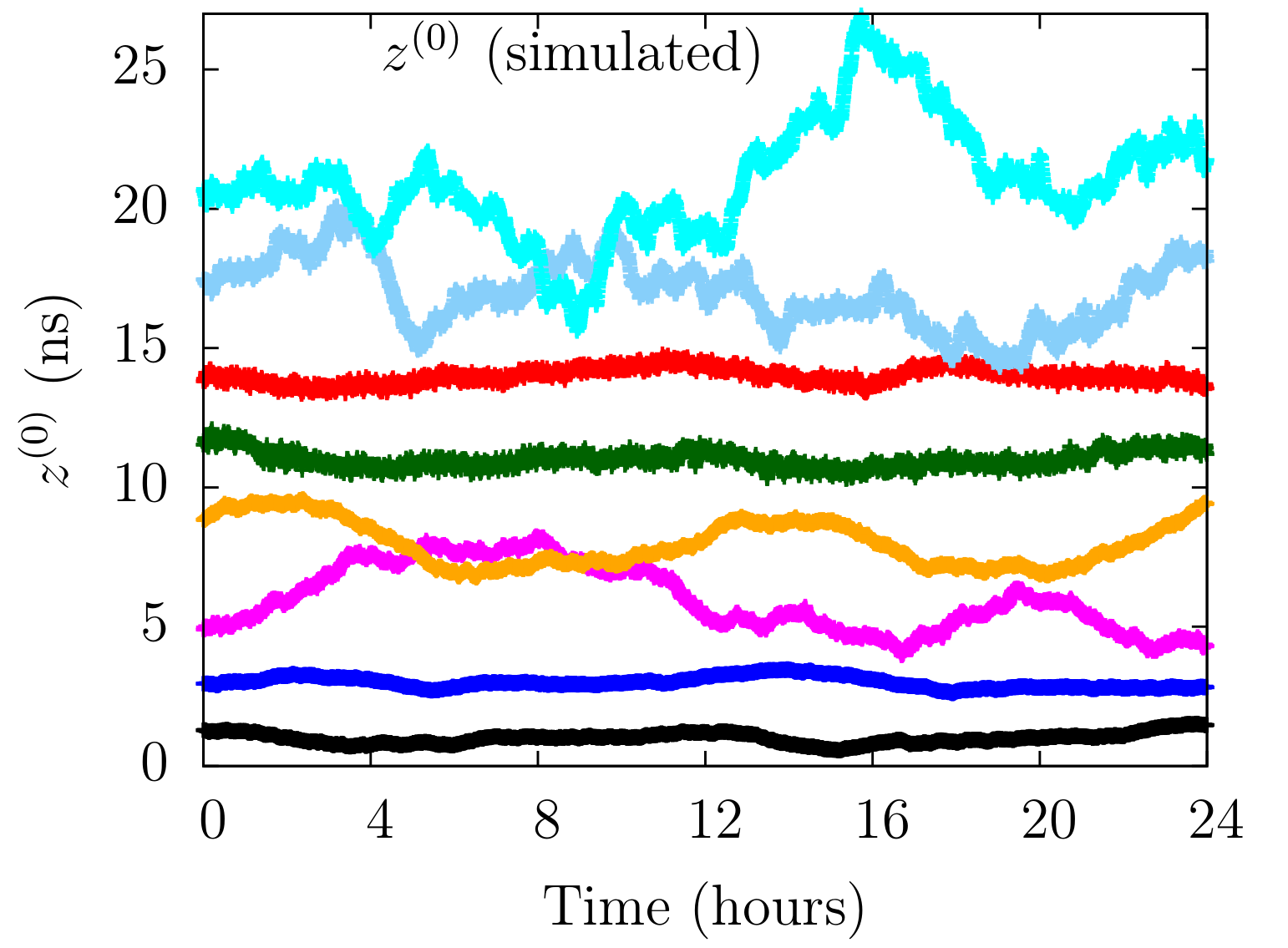}
	\caption{\small Comparison of (polynomial-reduced) real GPS clock data  for a few satellites from 21 June 2015 UTC (left) with simulated data for the corresponding SVNs (right). Each time-series is shifted by a constant offset for clarity. The curve labels encode the clock type, GPS block, and SVN.
	}
	\label{fig:simulated}
\end{figure*}
\begin{figure*}
	\includegraphics[width=0.435\textwidth]{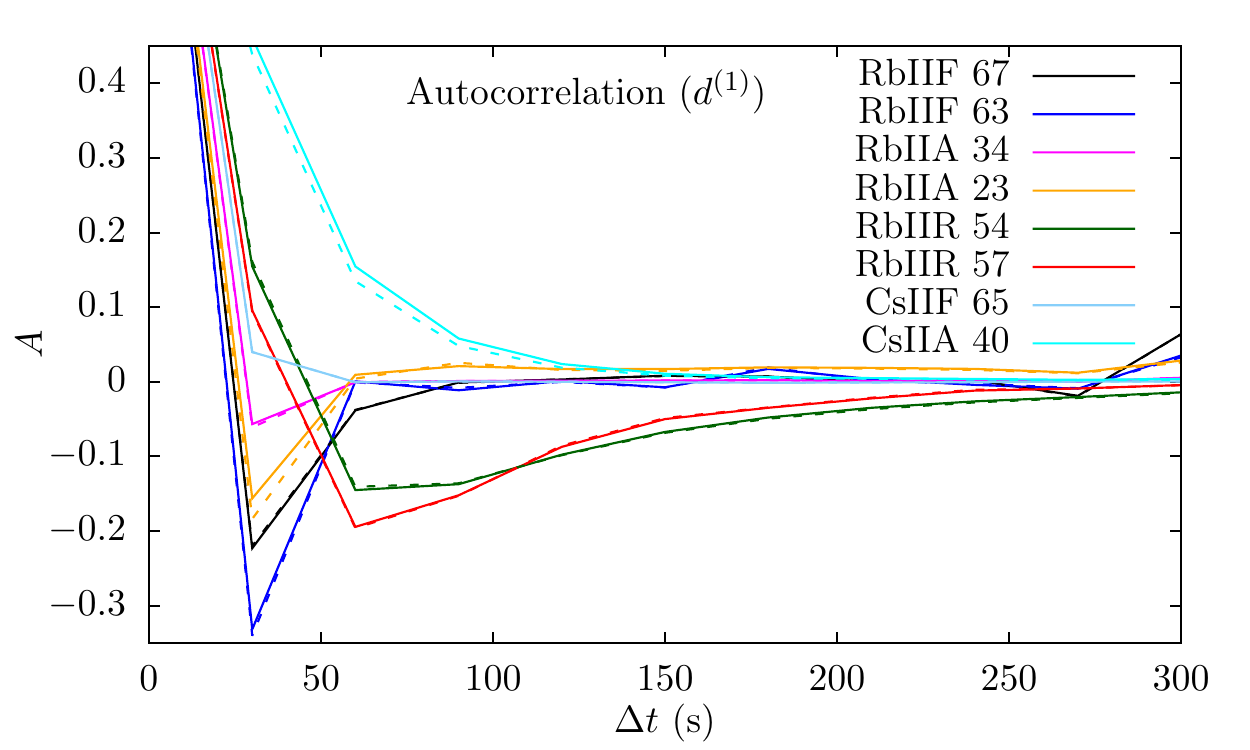}~~~~~
	\includegraphics[width=0.435\textwidth]{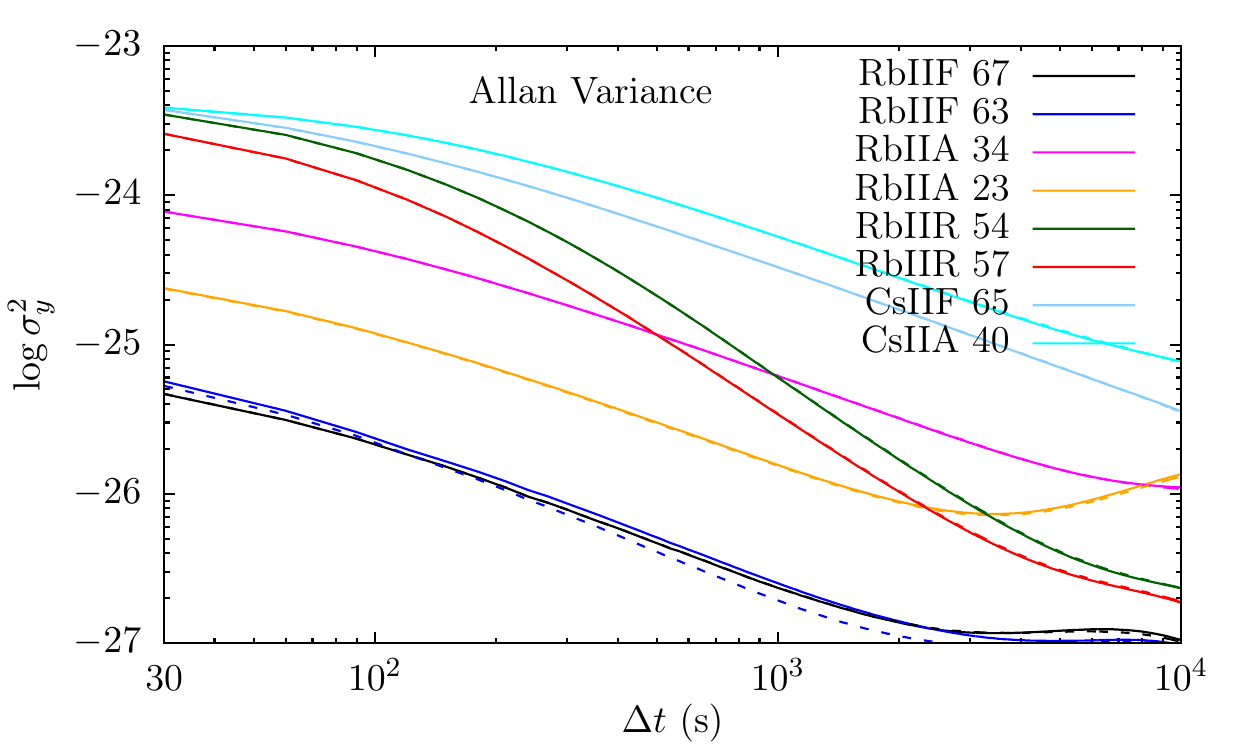}
	\caption{\small Comparison of the autocorrelation functions (left) and Allan variances (right) for the real data to those for the simulated data. Clocks are the same as in Fig.~\ref{fig:simulated}. The solid lines are from the real data, and the dotted lines are from the simulated data; they are practically indistinguishable.
}
	\label{fig:simulatedAVARACF}
\end{figure*}

We generate simulated time series data that have the same noise characteristics as the real clock data for each individual GPS satellite.
This is achieved by ``coloring'' pseudo-random white noise with the known power spectral densities for each clock~\cite{AlexThesis2016}.
We calculate the power spectral densities for each specific SVN using the clock data provided by JPL, as the clock performance may degrade over time, and the clocks can perform differently when in orbit than when tested in a laboratory environment. We also simulate cross-correlations (correlations between different clocks).
This is achieved by simulating a reference clock, which adds a common noise stream to all the clocks in the network.

In Fig.~\ref{fig:simulated}, we plot several arbitrarily selected real JPL clock solutions, $d^{(0)}$, alongside the simulated clock solutions for the corresponding SVNs (denoted $z^{(0)}$) to demonstrate the quality of simulated data.
The standard deviations of the simulated data (after first- or second-order differencing) match exactly those of the real data for the given SVNs.  
Further, the longer-scale noise characteristics also match  -- in Fig.~\ref{fig:simulatedAVARACF}, we plot the autocorrelation functions and Allan variances for both the simulated and real data for the same clocks. 
These figures demonstrate that the simulated clock data do indeed have the same noise characteristics as the real data.

Having generated simulated time series, we can test our Bayesian search code in a number of distinct ways:
\begin{enumerate}
\item
To gauge the prevalence of statistical false-positives, we run the code for the event-free simulated data.
\item
We inject DM event signals into the simulated data streams to gauge the efficacy of our technique to pick out true-positive events.
\item
We inject ``bad'' events (i.e., signals that are not properly correlated) into the simulated data as a test of the robustness of the search technique.
\item
We use parameter estimation to extract the observed parameters of the injected DM event, and compare the results to those used to generate the injected DM signal as a test of the method accuracy and efficacy.
\end{enumerate}

\subsection{Prevalence of statistical false-positives}

\begin{figure}
	\includegraphics[width=0.475\textwidth]{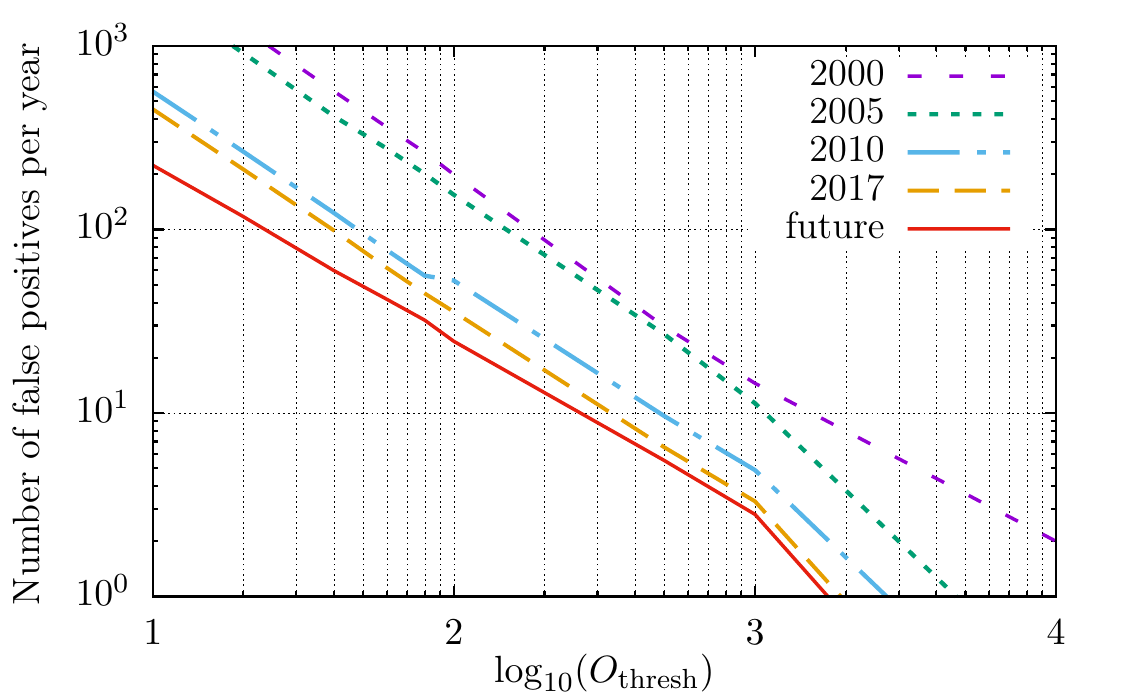}
\caption{\small 
	Rate of statistical false positives as a function of the odds ratio threshold, $O_\mathrm{thresh}$,  for thin domain walls. 
	The rate of false positives from simulated GPS networks typical for the given years: 
	2000 (1 Rb-II, 7 Rb-IIA, 3 Rb-IIR, 5 Cs-II, 11 Cs-IIA),
	2005 (1 Rb-II, 8 Rb-IIA, 12 Rb-IIR, 1 Cs-II, 8 Cs-IIA),
	2010 (5 Rb-IIA, 19 Rb-IIR, 2 Rb-IIF, 5 Cs-IIA, 2 Earth-based H-masers),
	2017 (19 Rb-IIR, 10 Rb-IIF, 5 Earth-based H-masers),
	and a possible future network (30 Rb-IIF--style satellites, 20 Earth-based H-masers).
	Each curve corresponds to 4 years of 30\,s sampled simulated data.
}
	\label{fig:FalsePos}
\end{figure}

We wish to define a threshold for the odds ratio, $O_{M,\bar{M}}(j_0)$, Eq.(\ref{eq:odds}). 
If the spike in the odds ratio is larger than this threshold, such an event can be investigated as a potential DM event.
In order to do this, we need to calibrate the rate of statistical false positives. 
To this end, we ran multiple simulations of various GPS clock network configurations, and computed the odds ratio (\ref{eq:odds}) for each epoch.
For each combination of clocks, we considered 2048 realizations of 2048 30\un{s}-epochs, amounting to approximately 4 years of simulated data for each simulation.
A plot of the rate of false positives  as a function of the threshold is presented in Fig.~\ref{fig:FalsePos}.
This plot is for the specific DM model of ``thin'' ($d \ll 10^4\un{km}$, see Appendix~\ref{sec:specific-signals}) domain walls.

Note that for this exercise, a false positive is counted whenever an epoch has a value for the odds ratio above the given threshold.
This is a conservative definition, since the ``width'' of the odds-ratio spike (due to the imperfect resolution) may lead to the same false-positive event appearing in more than one neighboring epoch. 
By our definition, this will be counted several times.

We can also drastically reduce the number of false positives that occur by introducing a minimum value (magnitude),  $h_{\rm min}$, for the integral over signal magnitudes (\ref{eq:m-likelihood}).
Of course, this also means we can only detect positive events with $|h|>h_{\rm min}$.
We can then perform the analysis in several sweeps, systematically reducing $h_{\rm min}$ each time until signals of a given magnitude can no longer be excluded.

\subsection{Detecting injected DM events}

\begin{figure}
	\includegraphics[width=0.425\textwidth]{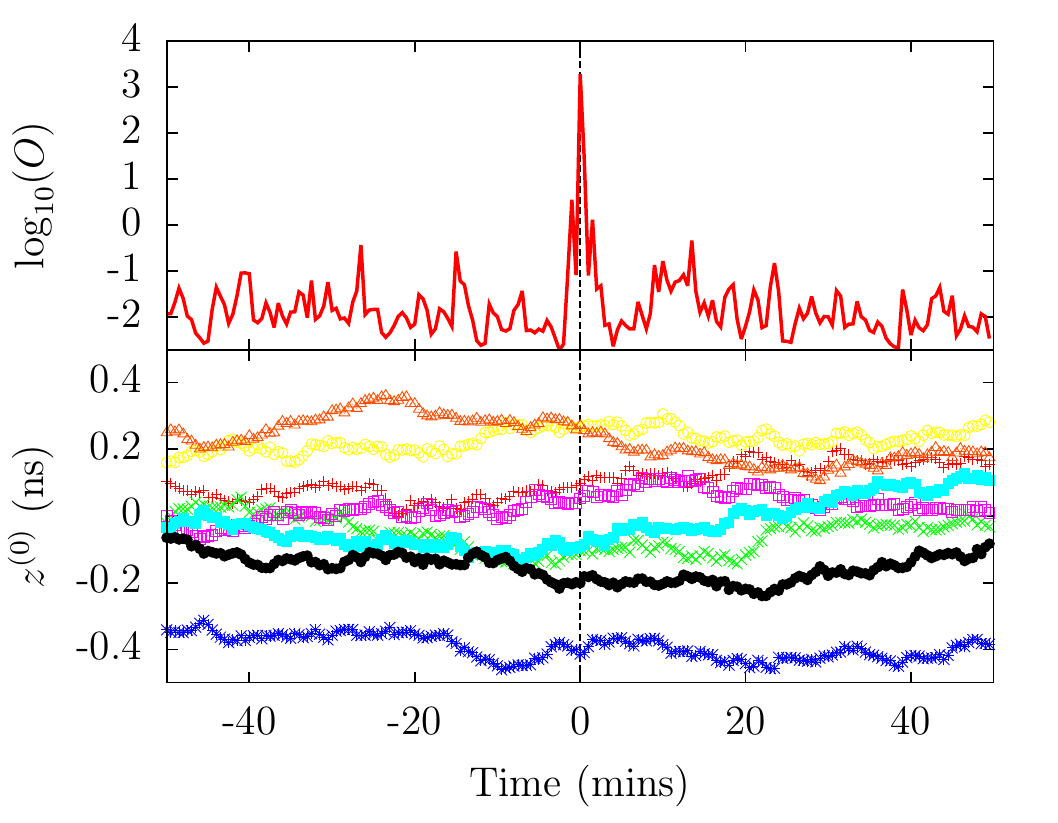}
	\caption{\small Bayesian detection of an injected thick domain wall ($d = 10^4\un{km}$) signal. The wall sweeps the GPS network of 32 satellite clocks (with $\s=0.01\un{ns}$) at time $t_0 = 0$.
For this simulation, $h = 0.02\un{ns}$. 
Bottom panel: simulated clock biases shown for the first 8 clocks (including the injected thin-wall signal). Each time-series is shifted by a constant offset for clarity. 
Top panel: the corresponding odds ratio using the same time scale.
}
	\label{fig:O}
\end{figure}

To determine the sensitivity of the method, we must know the probability of positively detecting DM events of a given magnitude.
To this end, we generate clock data per Sec.~\ref{sec:TestMethod-randts}, inject randomized DM signals into the data streams and compute the odds ratios. In Fig.~\ref{fig:O}, we present one such simulation as an example. 
Here we show the first 8 (of 30) simulated time series' for a 1.5\,hr window.
In this example, for simplicity, the clock noise was taken to be white (in $d^{(1)}$).
Then we injected a single ``thick'' domain-wall event for a wall of size $d=10^{4}\, \mathrm{km}$; the velocity and incident direction were chosen randomly.
The odds ratio was calculated for each epoch.
The spike in the odds ratio at the event is apparent, while the event is not discernible by eye in the data streams.
Note that the search routine is isolated from the simulation -- it is not made aware of the event time, speed, direction, magnitude, wall width (or if there was an event at all).

\begin{figure}
\includegraphics[width=0.475\textwidth]{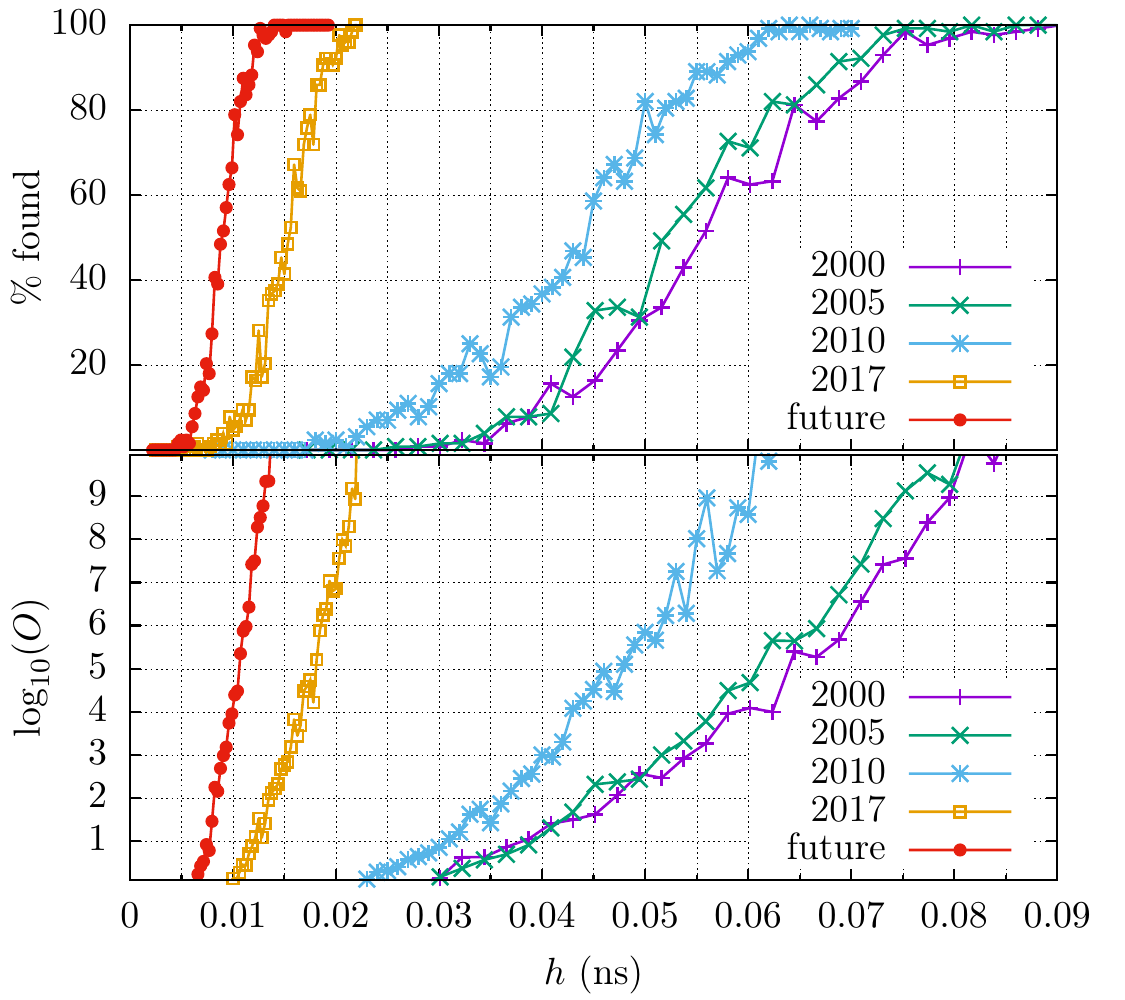}
\caption{\small 
	Efficacy of the method for detecting injected thin-wall DM signals, for the same simulated networks as in Fig.~\ref{fig:FalsePos}. Each point represents 128 trials, each curve has $\sim60$ points.
	Top panel shows the fraction of injected thin-wall events that were correctly identified, as a function of the event magnitude $h$.
	The odds threshold was set to allow fewer than $10$ false positives per year (see Fig.~\ref{fig:FalsePos}).
	Bottom panel shows the average log-odds ratio  as a function of $h$ on the same scale. 
}
	\label{fig:TruePos}\label{fig:tp-odds}
\end{figure}

Figure~\ref{fig:TruePos} shows the fraction of injected thin domain wall events that are correctly identified, as a function of the signal magnitude.
The velocity and incident direction for each wall was chosen randomly (according to the SHM distributions, Fig.~\ref{fig:distros}), and we assumed all clocks were affected by the DM in the same way (i.e., all clocks have the same $\Gamma_{\rm eff}$).
For this analysis, the odds threshold was set to allow fewer than 10 false positive events per year ($O_\mathrm{thresh} \sim 10^3$, see Fig.~\ref{fig:FalsePos}).
Note, for 30\,s data, there are over $10^6$ epochs in a year.
We count an event as ``found'' if there was a spike in the odds ratio above the determined threshold that appears within $\pm1$ epoch from the injected incident time $t_0$.
(Of course, occurrences where a single event leads to an odds-ratio spike for more than one epoch are not double-counted, only one event is injected per trial, and it is either found or not.)
Also shown in Fig.~\ref{fig:tp-odds} is the average of the log odds ratio for each of these simulated networks as a function of the magnitude of the injected domain wall signal.
The large ``gap'' in the sensitivity that occurs around 2010 is due to the introduction of the Rb-IIF satellite clocks, which are substantially more stable than the older generation satellite clocks; see Appendix~\ref{sec:clocknoise}.

In Fig.~\ref{fig:TruePos-white}, we show the same true- and false-positive test results, but for networks of a varying number of identical pure white frequency noise devices ($d^{(1)}$ equivalent).
This is to demonstrate the general efficacy of the method, without specific reference to the properties of the GPS data.

Note that the ground receiver clocks contribute only minimally, even though they are significantly more precise than the GPS satellite clocks.
That is because our current data is sampled only every 30\,s, which is about the time it would take for a DM object to cross the Earth, meaning many of the Earth-bound clocks will be affected by the DM during the same data acquisition interval.
As discussed in Sec.~\ref{sec:noise}, it is possible to re-process the existing raw GPS data to produce 1\,s sampled time series.
In addition to the statistical improvement from the larger data set, this would also further allow us to take full advantage of the highly-stable Earth-based receiver and laboratory clocks.
Of course, this advantage comes at the cost of significantly increased computation time, which scales (roughly) quadratically with the number of data points $J_{W}$ due to the correlations, see Eq.~(\ref{eq:chis}).

\begin{figure}
\includegraphics[width=0.475\textwidth]{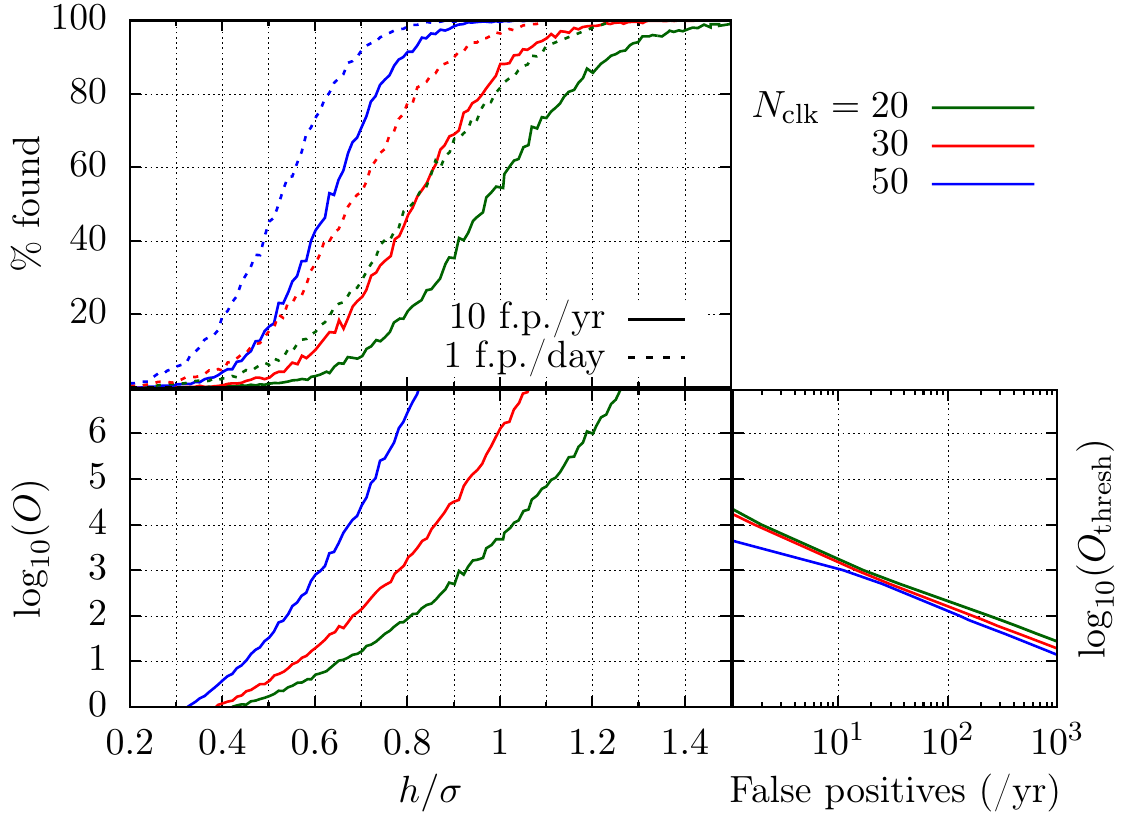}
\caption{\small 
Monte-Carlo simulations for thin domain walls, using a network of pure white-noise (in $d^{(1)}$) devices.
The green, red, and blue curves are for a network of 20, 30, and 50 identical devices, respectively. 
Top panel shows the fraction of events that were correctly identified, as a function of the injected event magnitude $h$ (scaled by $\s$, the standard deviation of the data noise).
This is done requiring an odds threshold such that there are fewer than 10 false positives per year (solid lines), or 1 false positive per day (dotted lines).
Bottom left panel shows the average log-odds ratio as a function of $h/\s$ on the same scale.
Bottom right panel shows the yearly rate of false positives as a function of the odds threshold, $O_{\rm thresh}$.
}
	\label{fig:TruePos-white}
\end{figure}

We also check the ``robustness'' of the method, to ensure incorrectly correlated events (that may exist in the data due to Earth-sourced or other non-galactic perturbations) are not flagged as potential events.
To do this, we inject a single perturbation of a specified magnitude into each satellite data stream at a random epoch, all within the same 5 minute time window.
This simulates a domain wall crossing, except in the important fact that the network perturbations are not correctly correlated between different satellites.
Injecting a large $2\sigma$ perturbation of this kind ($\sigma$ is the typical standard deviation of the clock noise) into the simulated data streams for a 30-clock network, fewer than 1\% present odds ratios anywhere within the 5 minute window that are above the threshold.

\subsection{Parameter estimation}\label{sec:parameters}

\begin{figure*}
	\includegraphics[width=0.32\textwidth]{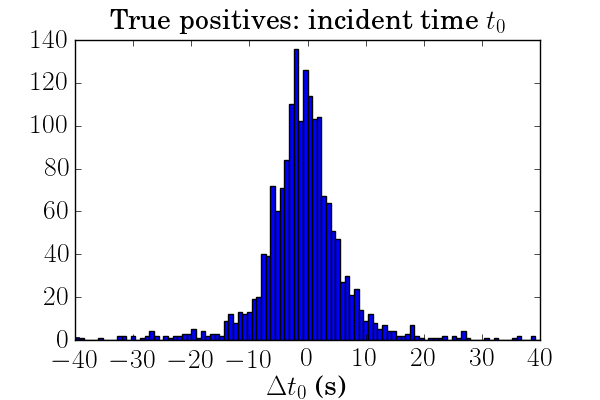}~
	\includegraphics[width=0.32\textwidth]{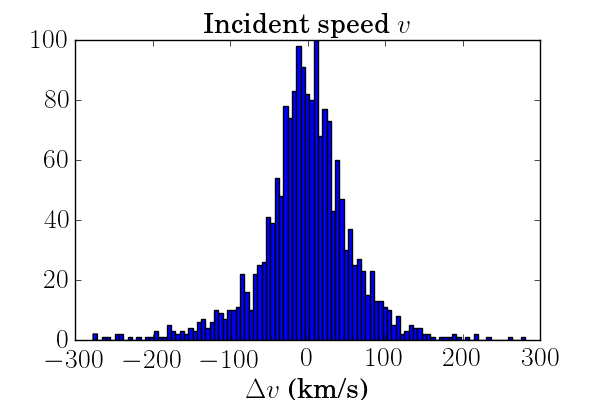}~
	\includegraphics[width=0.32\textwidth]{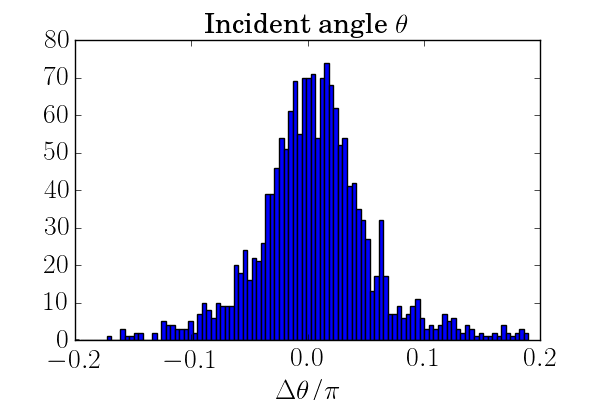}
	\caption{\small Example normalized histograms for the difference between the injected event parameters and the best-fit values extracted from the Bayesian analysis. Results for 2048 randomized simulations of a 25 satellite clock homogeneous network. Each  trial has a single $\sim1\s$ thin wall event injected with $v\simeq300\un{km}\un{s}^{-1}$.
	{\em Left}: for the incident arrival time, $t_0$, {\em middle}: for the scalar speed, $v$, and {\em right}: for the incident polar angle, $\theta$. 
	We have resolution of better than $\sim\pm0.1\pi$ radians for the incident angle, and $\sim\pm10\un{s}$ for the incident time (note that this is with 30\un{s} sampled data).}
	\label{fig:params}
\end{figure*}
\begin{figure*}
	\includegraphics[width=0.32\textwidth]{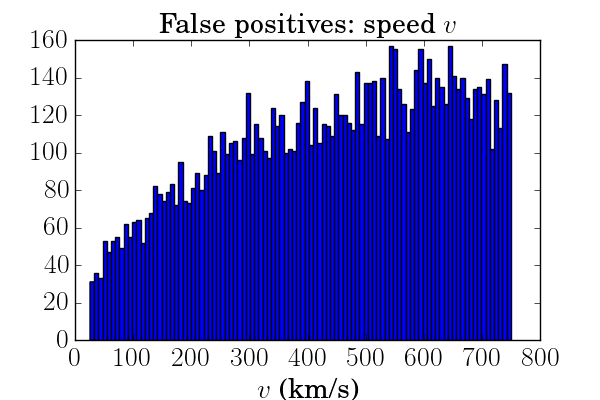}~
	\includegraphics[width=0.32\textwidth]{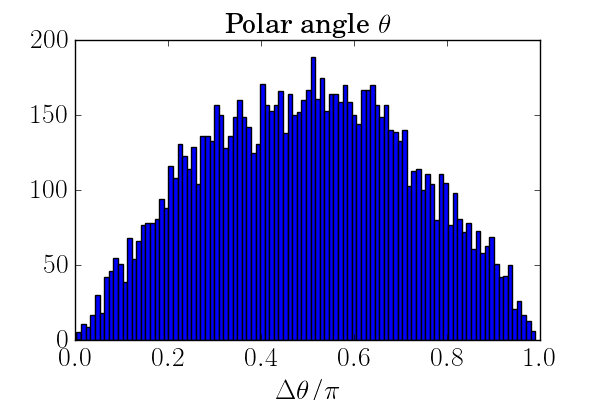}~
	\includegraphics[width=0.32\textwidth]{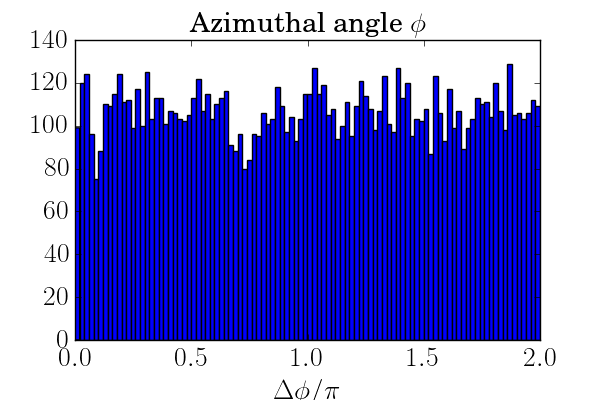}
	\caption{\small Normalized histograms for the distribution of the best-fit values extracted from the false positives of the Bayesian analysis, {\em left}: for the scalar speed $v$, {\em middle}: for the polar angle $\theta$, and {\em right}:  the azimuthal angle $\phi$. 
	(The hump in the $\theta$ histogram is due to the solid angle volume element $\sin\theta$.)
	Here, a low threshold ($O_{\rm thresh}=10$) was chosen to increase the statistics; when increasing $O_{\rm thresh}$, the shape of the histograms remains constant (it is prohibitively computationally intensive to run enough simulations to form false positive histograms for large $O_{\rm thresh}$, see Fig.~\ref{fig:FalsePos}).
For these simulations, the priors were excluded (i.e., flat priors were assumed).
	}
	\label{fig:fp-params}
\end{figure*}

When a spike in the odds ratio is  above the pre-determined threshold value, we can investigate this region of data as a potential event.
For example, by finding the set of ``best-fit'' parameters that maximize the un-marginalized likelihood, 
we can estimate the properties of the possible event (e.g., the time of arrival, size of the object, coupling strength etc.). 
In Fig.~\ref{fig:params}, we show histograms of the parameter estimation for a number of simulated trials where event signals were randomly injected into simulated data.
Shown in the plots is the difference between the injected value and the extracted best-fit value for the crossing time $t_0$, speed $v$, and incident polar angle $\theta$, for simulated domain wall crossings.
These parameters are representative of the spatial and temporal resolution of the method.
Note that Fig.~\ref{fig:params} was generated for 
 30\,s sampled data -- using the re-processed 1\,s data (as discussed above) will lead to a substantially improved resolution in the arrival time and velocity.

We also perform the parameter estimation for the false positive trials, where the analysis is performed on simulated event-free data.
The resultant histograms are presented in Fig.~\ref{fig:fp-params}.
In this case, when neglecting the priors, the histograms are flat, with a  slight bias of more false-positives towards higher velocities.
When including the priors, the distribution of extracted parameters from the false-positives match the priors, as expected.

This means that there is a potential to search for events even below the ``false positive floor''.
Reducing the odds ratio threshold will allow us to detect much smaller DM events, but will also lead to a larger number of false positives.
The true positive results, however, are expected to follow the distribution of velocities and incident directions predicted by the standard halo model.
This is relevant for the part of the parameter space with $\CT\ll1\un{year}$.
There would also be expected annual modulations in the event rate, average event speed, and most-common incident direction, see Sec.~\ref{sec:directional}.
In this case, the analysis would need to be performed without the priors (i.e., assuming flat priors) to avoid biasing the false positives.

\section{Search sensitivity and discovery reach}\label{sec:results}

Combining Eqs.~(\ref{eq:A2}), (\ref{eq:variation}), and (\ref{eq:generalsignal}), we find the maximum signal amplitude observable in a given clock ($a$, with reference clock $R$) for a  domain wall crossing to be
\begin{equation}\label{eq:s1max}
s^{(1)}_{\rm max} \simeq  (\hbar c)  \rho_{\rm DM} \sqrt{\pi} d \tilde\tau v_g \mathcal{T} \left[   \Gamma_{\rm eff}^a  -  \Gamma_{\rm eff}^R\exp\left(-\frac{L^2}{d^2}\right)  \right],
\end{equation}
where 
the interaction duration is given $\tilde\tau = d/v$ for $d/v<\tau_0$ and $\tilde\tau = \tau_0$ otherwise ($\tau_0=30\,$s is the time period between data sample points for GPS),
and $L\sim10^4\,{\rm km}$ is the  distance between the clock  and the reference clock.

\begin{figure*}
	\includegraphics[width=0.47\textwidth]{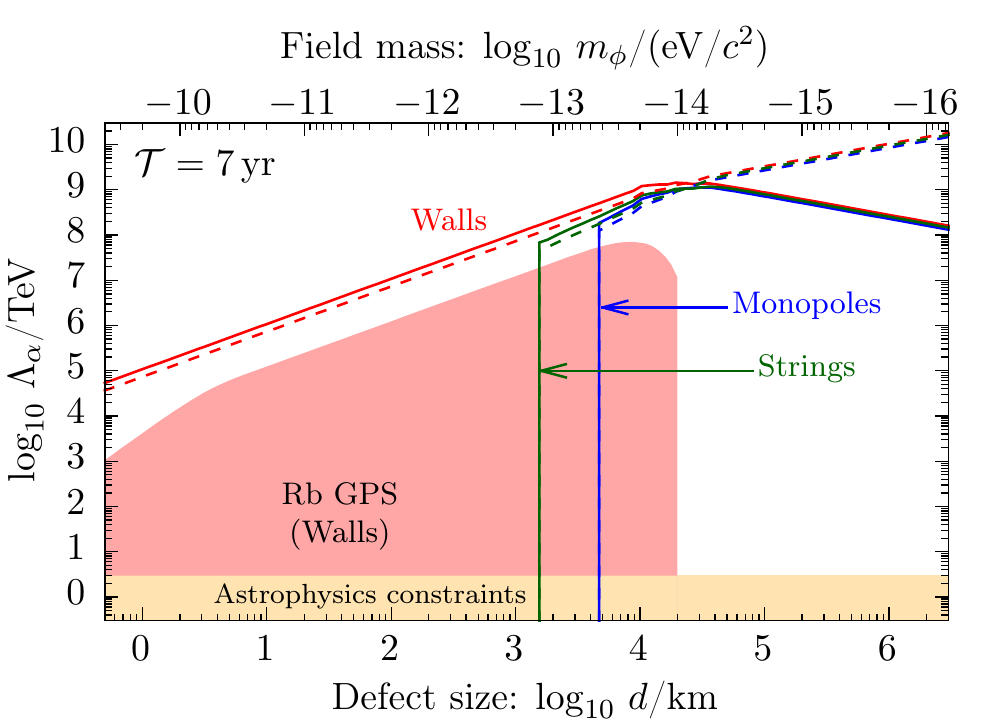}~~~
	\includegraphics[width=0.47\textwidth]{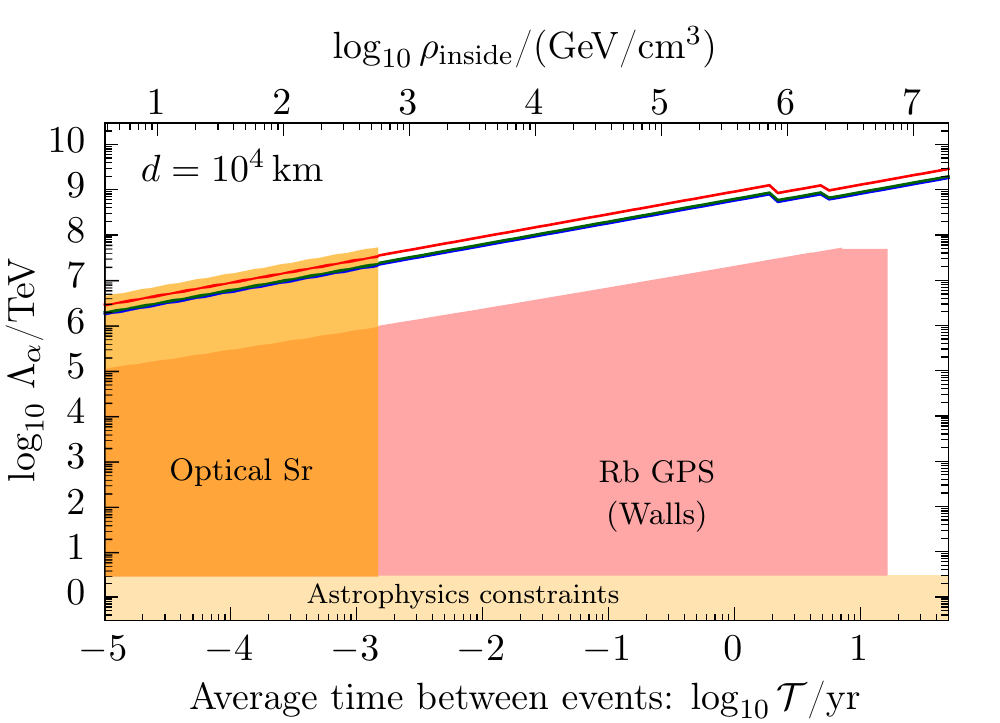}
	\caption{\small Projected discovery reach for topological defect dark matter, along with existing constraints for comparison. The red shaded region are the limits (on domain walls) from the initial GPS.DM search using the Rb GPS network~\cite{GPSDM2017}, the shaded orange regions are limits set by optical Sr clock~\cite{Wcislo2016} and from astrophysics observations~\cite{Olive2008}; these apply for walls, strings, and monopoles.
	The curves represent the projected sensitivities for our method, with the red, green, and blue colors for domain walls, strings, and monopole-like dark matter, respectively.
	For monopoles and strings, we require that at least 3 clocks are affected in the DM crossing, which causes the sharp cut-off for low $d$.
	The solid lines are the projections for the global network of GPS microwave clocks, and the dashed lines are the reach for the case when a single optical clock can be incorporated into the analysis.
	The sensitivity is slightly lower for large $\CT$, since we rely on the older GPS clocks.}
	\label{fig:proj}
\end{figure*}

The subtraction of two terms in square brackets in Eq.~(\ref{eq:s1max}) is due to the fact that when the maximum of DM field affects the clock, the reference clock is affected by its exponentially-suppressed tail.
When employing a network of identical clocks, this term leads to a fast decline in sensitivity for large $d$. 
This is because the clock and reference clock are affected in the same way, so no bias is built up between them.
In contrast, when employing clocks with significantly different effective couplings $\Gamma_{\rm eff}$ (particularly, combining microwave and optical clocks) this suppression is not realized.

Statistically, the minimum detectable signal is proportional to
\[
s^{(1)}_{\rm min} \propto \frac{\s_y(\t_0)\t_0}{\sqrt{N_{\rm clk} N_{\rm pts}}},
\]
where $N_{\rm clk}$ is the number of affected clocks, and $N_{\rm pts} \sim {d}/{v_g\t_0}$ is the number of data samples per clock for which the DM-induced signal is appreciable, and $\s_y(\t_{0})$ is the Allan deviation.
The proportionality constant depends on the efficacy of the search technique, and on $O_{\rm thresh}$, the odds ratio threshold required to eliminate false positives.
Therefore, we should have sensitivity to the region
\begin{equation} \label{eq:GammaX}
\Gamma_X \gtrsim \varepsilon \, \frac{\sigma_y(\tau_0)\tau_0 \, (\k_X^a-\k_X^Re^{-L^2/d^2})^{-1}}{\h c \rho_{\rm DM}\,\sqrt{N_{\rm clk} N_{\rm pts}}\,\tilde\t v_g\,  d \, \CT },
\end{equation}
where the factor $\varepsilon\sim O(1)$ is the efficiency factor determined from the simulations, and depends on $O_{\rm thresh}$.
From the results presented in Fig.~\ref{fig:tp-odds}, for a 90\% detection confidence level, and when requiring fewer than ten false-positives per year, we have $\varepsilon\approx5$ for the existing GPS data.
Future improvements in the search method should allow us to further decrease $\varepsilon$.

The $\varepsilon$ factor depends only fairly weakly on the search parameters.
For example, as shown in Figs.~\ref{fig:FalsePos} and Fig.~\ref{fig:TruePos-white}, increasing the odds threshold by a factor of 10 decreases the number of false positives by a factor of 10, while only increasing $s_{\rm min}$, the smallest detectable signal magnitude, by $\sim10\%$.
Therefore, we may estimate that for a 90\% detection confidence level, and when requiring fewer than one false-positive every 10 years,  $\varepsilon\approx6$.

The average time between consecutive encounters with a DM object, $\CT$, is considered a free parameter in our model (set by the number density of the DM objects).
The dependence of Eq.~(\ref{eq:GammaX}) on $\CT$ comes via the DM field amplitude (\ref{eq:A2}), and the requirement to not oversaturate the galactic DM density; the higher the number density of objects, the lower the field amplitude must be per object to compensate.
In order to determine the maximum $\CT$ that one can have sensitivity to, we assume the sequence of DM events can be modeled as a Poissonian process.
For example, if we expect one DM object to cross the Earth every period of $\CT$ on average, then in order to be $\sim90\%$ confident that an event would have occurred in the observation time $T_{\rm obs}$, we must require $T_{\rm obs} \gtrsim 2.3 \CT$.

We present the projected sensitivity of our search in Fig.~\ref{fig:proj}, along with the existing constraints.
To be consistent with existing literature, we present the sensitivity in terms of the effective energy scales,
$\Lambda_X \equiv 1/\sqrt{\abs{\Gamma_X}}$.
Specifically, we show the projections for $\Lambda_\alpha$; the projections for $\Lambda_{m_e}$ and $\Lambda_{m_q}$ are essentially the same, the only difference arising from the different sensitivity coefficients $\kappa_X$, see Eq.~(\ref{eq:Crb}).

The reduction in sensitivity above $d\simeq10^4\un{km}$ for the homogeneous clock networks is due to the fact that large DM objects will interact with the clock and reference clock at the same time, see Eq.~(\ref{eq:s1max}).
That is, above this value, we are only sensitive to the gradient in the DM field when using a homogeneous network.
The limits from our previous work \cite{GPSDM2017} have a sharp cut-off above this value, since in that work, we required that the DM signal would be present for just a single data point (see Appendix~\ref{sec:specific-signals}).
The Bayesian method presented in this work does not suffer this constraint.

Performing the simulations for strings and monopoles is substantially more computationally demanding, due to the number of extra free parameters that must be marginalized over (see Appendix~\ref{sec:specific-signals}).
However, the sensitivity can be approximated by analogy with the domain wall case.
For $d\gg R_{\rm GPS}$, the monopole and domain wall cases are essentially the same.
For $d < R_{\rm GPS}$, the sensitivity of the search can be estimated by noting the typical number of clocks that would be affected in a monopole crossing,
$N_{\rm eff} \simeq \left\lceil N_{\rm clk}   {d^2}/{R_{\rm GPS}^2}\right\rceil$.
A similar equation exists for strings, $N_{\rm eff} \propto   N_{\rm clk}   {d}/{R_{\rm GPS}}$.
For strings and monopoles, we required that at least 3 satellite clocks are affected during the DM sweep, $N_{\rm eff}\geq3$, which leads to a sharp drop in sensitivity for small $d$, as shown in Fig.~\ref{fig:proj}.

\section*{Conclusion}
We have described a method to use data from a distributed global network of precision measurement devices to search for transient signals that may be associated with sweeps by macroscopic-scale dark matter.
In particular, we considered the network of microwave atomic clocks on board the GPS satellites and ground stations, for which nearly two decades of archival data is available.
The method was demonstrated using simulated atomic clock data, and the prospects and discovery reach for topological defect dark matter was presented.
This approach can be extended in a straightforward fashion to other networks of high-precision measurement devices.

\acknowledgements
This work was supported by the U.S. National Science Foundation grant PHY-1506424.
We thank Chris Pankow, Derek Jackson Kimball, and Tigran Kalaydzhyan for discussions.
BMR is grateful to the CIERA institute and Northwestern University for hospitality during the CIERA Data Analysis Workshop,
and to the Perimeter Institute for Theoretical physics for support to attend the New Directions in Dark Matter and Neutrino Physics workshop and acknowledges the many helpful discussions that took place there.


\begin{widetext}
\section*{Appendix}
\end{widetext}
\appendix
\section{Velocity distribution and event rate}\label{sec:vel-distro}

Assuming the standard halo model, the velocity distribution of DM objects in the galactic rest frame is isotropic and quasi-Maxwellian, with dispersion of $290\un{km}\un{s}^{-1}$ and a threshold above the galactic escape velocity of  $v_{\rm ge}\simeq 544\un{km}\un{s}^{-1}$.
The vector velocity distribution for DM objects that cross paths with the Earth can be expressed in the Earth-centered inertial (ECI) frame as 
\begin{equation}
f_{\vv{v}}(\v{v})= C v \exp\left[- \frac{(\v{v}+\v{v}_g)^2}{v_c^2}\right]\Theta(v_{\rm e}(\psi)-v),
\label{eq:vector-v}
\end{equation}
where $\Theta$ is the Heaviside step function with
\[v_{\rm e}(\psi)=\sqrt{v_{\rm ge}^2-v_g^2\sin^2\psi}-v_g\cos\psi\]
being the effective escape velocity (the maximum allowable relative DM velocity as a function of $\psi$),
$\psi$ is the angle between $\v{\hat n}_{\rm g}=\v{v}_{\rm g}/v_g$ (the direction of Earth's motion through the galaxy) and $\v{\hat n}=-\v{v}/v$ (the vector of the incident DM object, $\cos\psi=-\v{\hat n}\cdot\v{\hat n}_{\rm g}$), 
$v_c$ is the speed of the Sun in the galactic rest frame, 
$v_g$ is the galactic speed of the local reference frame (ECI),
and $C$ is a normalization constant.
For the purposes of this work, we can neglect the smaller relative velocity of the Earth in its orbit around the Sun, and take
$v_g\approx v_c\approx 220\,{\rm km/s}$.
For a more detailed overview, see, e.g., Ref.~\cite{Freese2013}.

The direction of motion of the solar system through the galaxy points towards the Cygnus constellation; in the ECI frame $\v{\hat n}_{\rm g}\approx(0.46,-0.49,0.74)^{\rm T}$.
The angular distribution function for events is obtained by integrating over  velocities
\begin{equation}
f_\psi(\psi) = 2\pi \int_0^\infty f_{\vv{v}}(v,\psi)\,v^2\sin\psi \, \d v .  
\end{equation}
In fact, more than 90\% of events would be expected to come from the forward-facing hemisphere, see Fig.~\ref{fig:distros}.
Similarly, the scalar velocity distribution can be found by integrating over the angles
\begin{multline}
f_v(v)=2\pi \int_0^\pi f_{\vv{v}}(v,\psi)\,\,v^2 \sin\psi\, \d \psi  \\
\approx C v^2\left[ \exp\left(\frac{-(v-v_c)^2}{v_c^2}\right) - \exp\left(\frac{-(v+v_c)^2}{v_c^2}\right)\right].
\end{multline}

Since we are focusing on macroscopic DM objects, it is also instructive to constructive to consider the distribution of crossing durations.
Define $\t=d/v$ to be the time it takes a DM object of width $d$ to pass by a point in space (similarly we can consider $\t_{\rm GPS}=D_{\rm GPS}/v$, time to sweep the entire GPS constellation by the DM object central point).
It is convenient to define the inverse velocity $u\equiv 1/v$, and its corresponding distribution function
\[f_u(u)\equiv\frac{ \d P_u(u)}{ \d u}=\frac{ \d v}{ \d u}\frac{ \d P_v(v)}{ \d v}=v^2f_v(v),\]
where $ \d P_u(u)$ is the infinitesimal probability for the DM object to have inverse velocity $u$.
Then, the distribution for the crossing times is
\begin{equation}
f_\t(\t)\equiv\frac{ \d P_\t(\t)}{ \d \t}=\frac{ \d u}{ \d \t}\frac{ \d P_u(u)}{ \d u}=\frac{(d/\tau)^2}{d}f_v(d/\tau).
\end{equation}

In the case of domain walls, we are actually interested in the distribution of perpendicular velocities $\v{v}_\perp$, see Fig.~\ref{fig:vperp}.
Note that the Earth is more likely to cross paths with walls that have velocities close to the normal (such objects sweep out a greater volume per unit time).
If $\eta$ is the angle between $\v{v}$ and $\v{v}_\perp$, see Eq.~(\ref{eq:eta}), then 
the probability of encountering a wall with this angle is proportional to $\cos\eta$, and can be expressed as
\begin{equation}
 \d P_\eta(\cos\eta)=2\cos\eta\,\d(\cos\eta) = 2 \frac{v_\perp}{v^2}\d v_\perp.
 \end{equation}
Therefore, we have $f_{\vv{v}_\perp}(\v{v}_\perp)\, \d v_{\perp}\equiv \d P_{\v{v}_\perp}(\v{v}_\perp)$, with 
\[
\d P_{\v{v}_\perp}(\v{v}_\perp)=\int\limits_{v_\perp}^{\infty}\int\limits_{-{\pi}/{2}}^{{\pi}/{2}} \, \d P_\eta(\cos\eta)\, \d P_{\vv{v}}(\v{v})\,\d v\,\d\eta ,
\]
which implies
\begin{equation}
f_{\vv{v}_\perp}(\v{v}_\perp)=2 \int\limits_{v_\perp}^\infty\int\limits_{-{\pi}/{2}}^{{\pi}/{2}}  \, 
{f_{\vv{v}}(\v{v})}
\frac{v_\perp}{v^2} \,\d v\,\d\eta.
\end{equation}
We can further find the angular, scalar, and crossing-time distributions as above, which are also presented in Fig.~\ref{fig:distros}.

\section{Clock noise profiles}\label{sec:clocknoise}

Here we present a brief overview of the noise characteristics of the GPS satellite clocks.
For more detail, including the analysis for each individual SVN, see the Supplementary Information.

\begin{table}%
\caption{\small Typical standard deviations for the first- and second-order differenced data (30\un{s} sampling time interval) for GPS satellite clocks. For individual SVNs, including the daily-variation uncertainty and how they vary over time, see the Supplementary Information.}%
\begin{ruledtabular}%
  \begin{tabular}{ll D{.}{.}{1.5} D{.}{.}{1.5}}%
\multicolumn{1}{c}{Clock}&\multicolumn{1}{c}{Block}&\multicolumn{1}{c}{$\sigma^{(1)}/{\rm ns}$}  &\multicolumn{1}{c}{$\sigma^{(2)}/{\rm ns}$} \\
\hline%
Rb	& IIF 		& 0.013	& 0.	021	\\
 	& IIR 		& 0.074	&  0.099	\\
 	& IIA 		& 0.040	&  0.059	\\
 	& II 			& 0.048	&  0.069	\\
Cs	& IIF 		& 0.087	&  0.121	\\ 
 	& IIA 		& 0.089	&  0.090	\\
 	& II 			& 0.083	&  0.071	\\
  \end{tabular}%
\end{ruledtabular}%
   \label{tab:clocknoise}%
\end{table}%

\begin{figure}
	\includegraphics[width=0.42\textwidth]{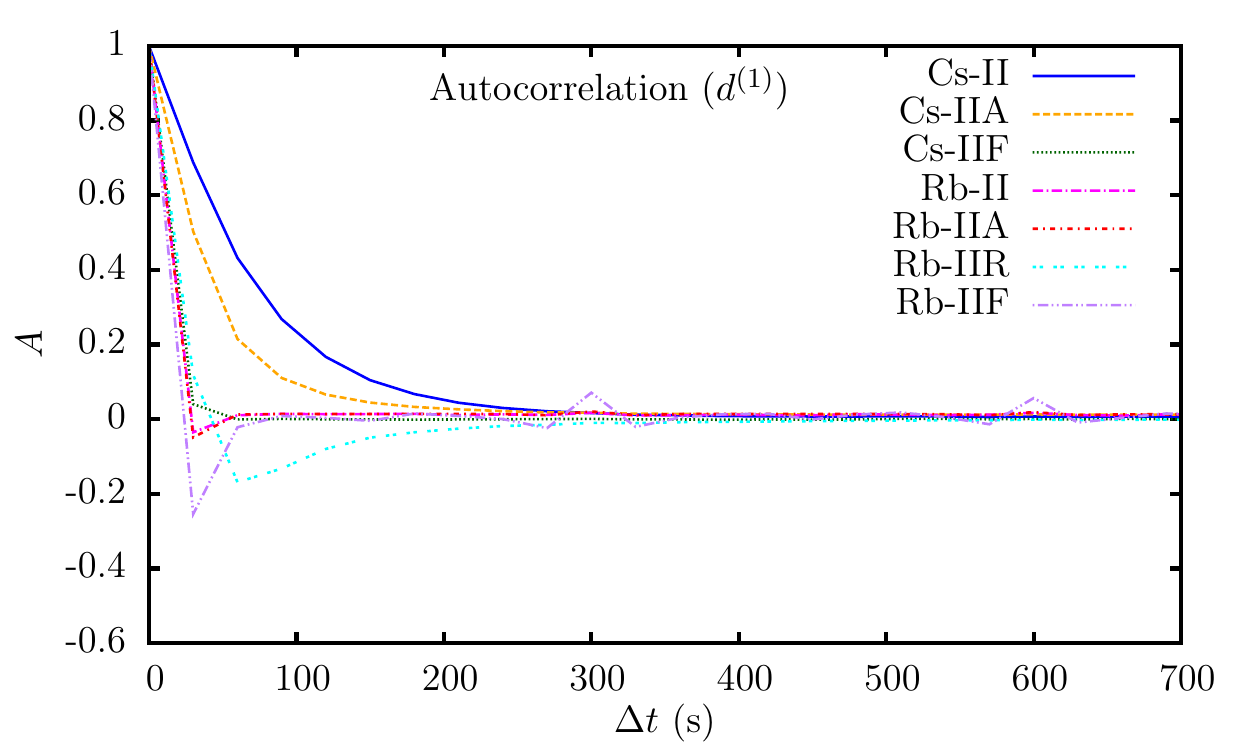}\\
	\includegraphics[width=0.42\textwidth]{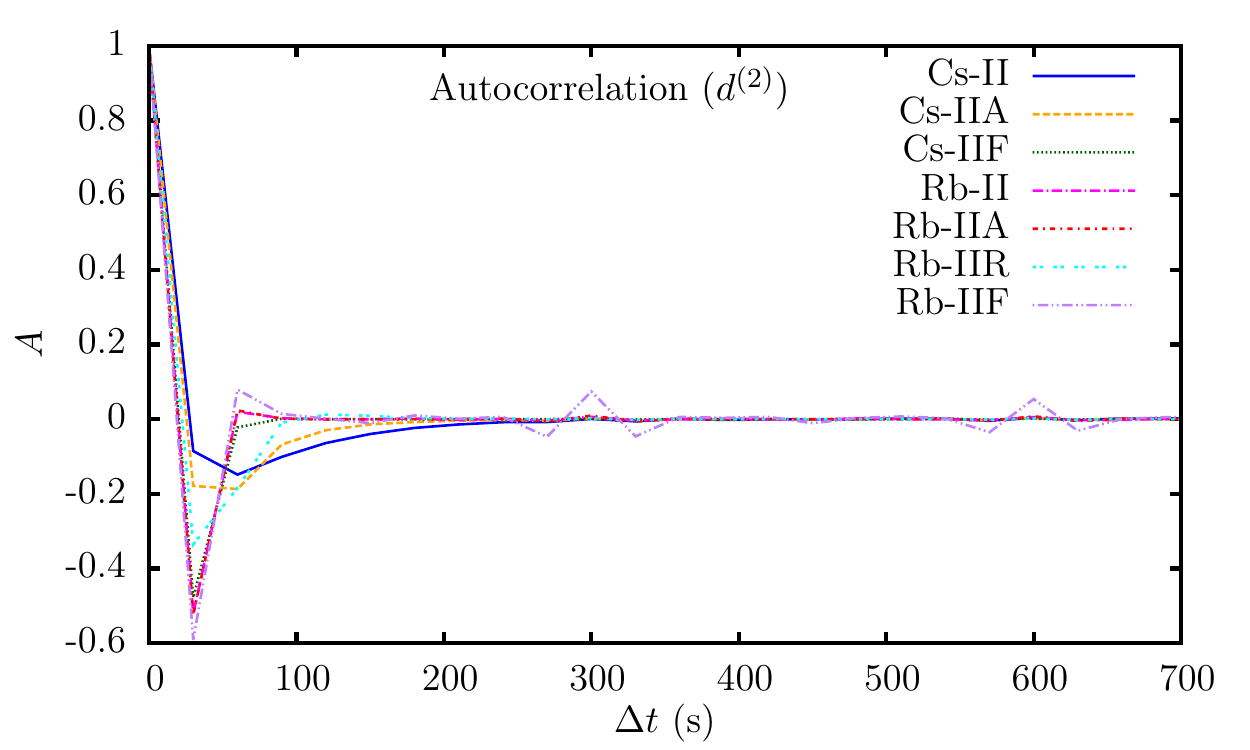}
	\caption{\small Averaged autocorrelation functions for first-order (top) and second-order (bottom) differenced clock data.
}
	\label{fig:ACF}
\end{figure}

In Table~\ref{tab:clocknoise}, we present the average standard deviations of each clock and satellite combination for both first- and second-order differenced data, averaged over all available SVNs and reference clocks.
We also form the autocorrelation function (ACF),
\begin{equation}
\label{eq:ACF}
	A^{a}(j \t_0)=\sum_{l=0}^{J-j-1} \frac{{d^a_l} ~ d^a_{l+j}}{(J-j)\,{(\sigma^a)}^2},
\end{equation}
for each clock, where the time-series data $\{{d^a_j}\}$ is assumed to be centered around 0, 
and $J$ is the total number of data points for each clock per day. 
For the 30\un{s} sampled data, $\t_0=30\un{s}$ and $J=2880$.
(Here, ${\sigma^a}$ is the standard deviation of the clock data, not the formal error.)
We calculate ACFs for the first- and second-order differenced data ($d^{(1)}$ and $d^{(2)}$).
In Fig.~\ref{fig:ACF} we show the ACF averaged over all clocks of a specific type between July 2004 and June 2016 for first- and second-order differenced data.
For pure white data $A(0)=1$ and $A(\t)\to0$ for $\t\neq0$; other noise profiles have distinct ACF forms (see, e.g., Ref.~\cite{Barnes1971}).
First-order differencing is sufficient to ensure all the Rb and the block IIF Cs clock time-series are sufficiently stationary, while the block II and IIA Cs clocks require second-order differencing.

\begin{figure}
	\includegraphics[width=0.42\textwidth]{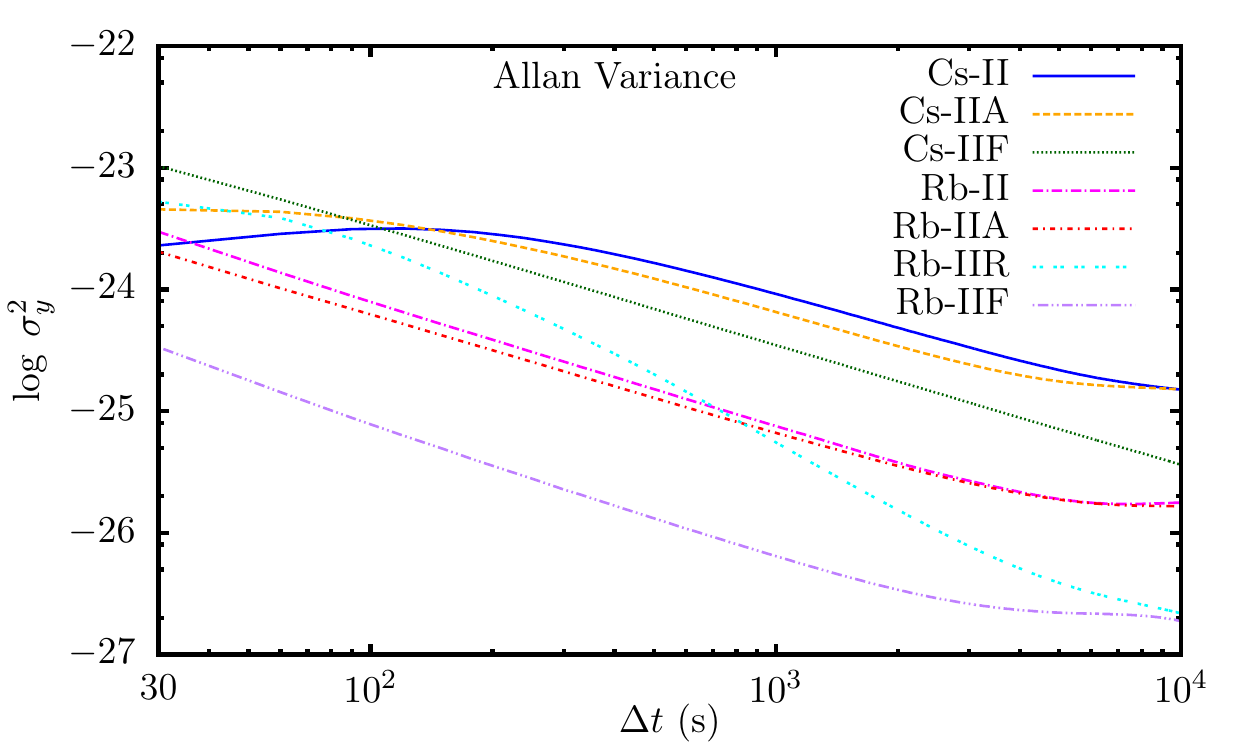}
	\caption{\small Allan variance (\ref{eq:AVAR}) for each GPS satellite block, averaged over all available SVNs. See also the Supplementary Information.}
	\label{fig:AVAR}
\end{figure}

We can also compute the Allan variance for each clock 
\begin{equation}
\label{eq:AVAR}
	\sigma_{y}^2(\Delta t)=\sum_{j=0}^{J-2l-1}\frac{\left( d_{j}-2d_{j+l}+d_{j+2l} \right)^2}
					    {2\,l^2\t_0^2\left(J-2l\right)},
\end{equation}
which is a widely utilized time-domain measure of the frequency stability \cite{Barnes1971}.
Note that the Allan variance is a function of the averaging time, which in our case can be written $\Delta t=l \t_0$.
The Allan variance is shown is Fig.~\ref{fig:AVAR}, where we use the non-differenced data, $d^{(0)}$.

\begin{figure}
	\includegraphics[width=0.47\textwidth]{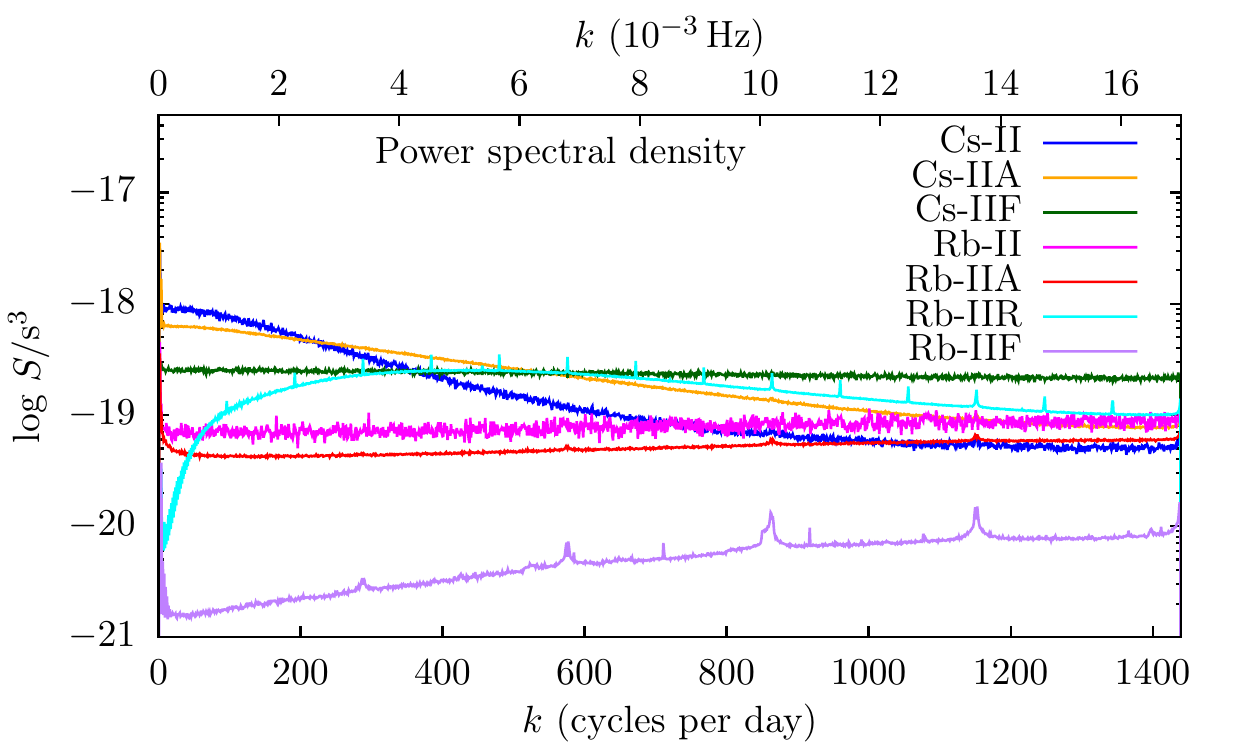}
	\caption{\small The averaged power spectral densities (for $d^{(1)}$) for the various clock/satellite block combinations. Note that this includes noise from the H-maser reference clock. See also the Supplementary Information.}
	\label{fig:PSD}
\end{figure}

For a given clock, $a$, we can form the power spectral density (PSD),
\begin{equation}
\label{eq:PSD}
	S^{a}(k)=\frac{\t_0}{J}\abs{\widetilde {d}^a(k)}^2,
\end{equation}
where 
$\widetilde d^a(k)=\sum_j d^a_{j} \exp\left({-i2\pi{ j k}/{J}}\right)$
 is the discreet Fourier transform  of the time-series data for the clock $a$. 
The PSD units are ${\rm s}^2/\un{\rm Hz}$.
Plots of the PSD for each of the clock/satellite combinations are shown in Fig.~\ref{fig:PSD}, for which we use the singly-differenced data, $d^{(1)}$.


The periodic spikes that appear in the power spectrum and autocorrelation function (particularly visible for the Rb-IIF satellite clocks, see Figs.~\ref{fig:ACF} and \ref{fig:PSD}) correspond to a 5-minute period, and are technical artefacts traceable to the partitioning method used in the initial JPL data processing.
This has been addressed in recent updates to their software.

\section{Specific dark-matter signals}\label{sec:specific-signals}

Here, we present the specific DM profiles and resultant signals for domain walls, monopoles, and strings.
We then link the general $h$ parameters back to the specific field parameters for those models.

\paragraph*{Thin walls---}

The simplest case to consider is thin domain walls.
By ``thin'', we mean that the width of the wall is sufficiently small such that it will pass through any clock within the 30\,s sampling period,
\[d\ll \sim 300\un{km}\un{s}^{-1}\times 30\un{s} \approx 10^4 \un{km}.\]
In this case, the profile can be considered to be a delta-function
$\varphi^2(t^a,t)=\delta(t^a-t),$
so that
\begin{equation}
	{s^a_j}^{(0)}=
	\begin{cases}
		0 		& t \leq t^a,\;t^R  \\
		h^a 		& t^a \leq t < t^R \\
		-h^R 	& t^a > t \geq t^R \\
		h^a-h^R	& t\geq t^a,\;t^R \\
	\end{cases}
\label{eq:sij-thinwall}
\end{equation}
($t=j\t_0$).
From the normalization defined in Eq.~(\ref{eq:generalsignal}), in the thin wall case, the parameter $h$ can be linked back to the field parameters as
\begin{equation}
\label{eq:h-thinwall}
h = A^2\sum_X{\k_X}\,{\Gamma_X}.
\end{equation}

\paragraph*{Gaussian profile walls---}

For walls of finite thickness, we assume a Gaussian density profile, with root-mean-square width $d$, such that
\begin{equation}
\varphi^2(t^a,t')=\frac{v_\perp}{d \sqrt{\pi}}\exp\left( -\frac{v_\perp^2}{d^2}(t^a-t')^2 \right).
\label{eq:profile-gaussianwall}
\end{equation}
The normalization coefficient, which includes $v_\perp$ and $d$, is chosen purely for convenience so that the maximum accumulated clock bias will be $h$, in order to be consistent with the thin wall case and because it is $h$ that is the directly observable parameter.
Then, the integral in Eq.~(\ref{eq:generalsignal}) can be expressed in terms of error functions, 
\begin{multline}
{s^a_j}^{(0)}=
\frac{1}{2}\Bigg\{ h^a - h^R \,+  \,h^a \erf\left[ \frac{v_\perp(j-t^a)}{d} \right]
\\
-h^R \erf\left[ \frac{v_\perp(j-t^R)}{d} \right] \Bigg\}.
\label{eq:sij-dwall}
\end{multline}
In the Gaussian-profile wall case, the parameter $h$ relates to the field parameters as
\begin{equation}
\label{eq:h-dwall}
h = \frac{A^2 d \sqrt{\pi}}{v_\perp}
\sum_X{\k_X}\,{\Gamma_X}.
\end{equation}

\paragraph*{Monopoles---}

\begin{figure}
	\includegraphics[width=0.45\textwidth]{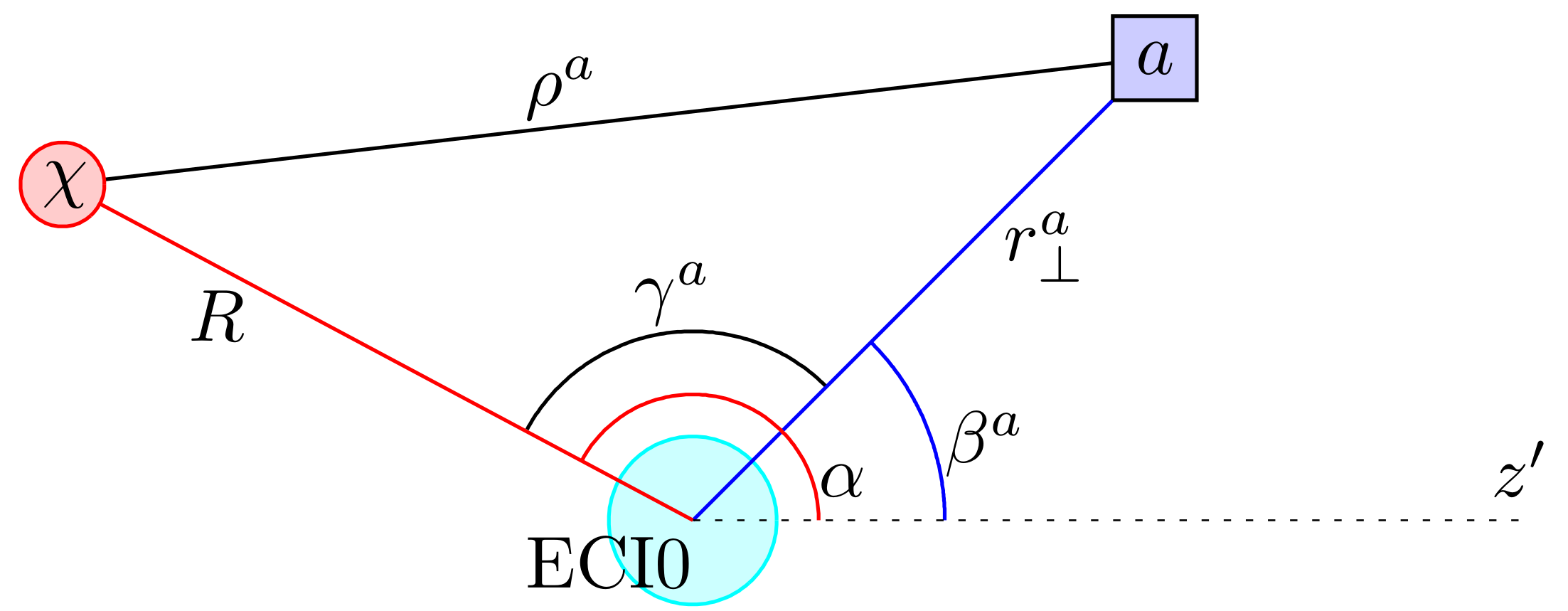}
	\caption{\small Geometry of a monopole object crossing the GPS constellation.
	The monopole (labeled $\chi$) enters along unit vector $\v{\hat n}$ (which points into the page), at perpendicular distance $R$ from ECI0 (the Earth center), and makes an angle $\alpha$ with respect to the ${\hat z}'$-axis in the plane perpendicular to $\v{\hat n}$;	${\rho^a}$  is the impact parameter for satellite $a$.}
	\label{fig:monopolegeom}
\end{figure}

For monopoles, we assume Gaussian profile spherical objects, and also have to consider the impact parameter, $\rho$, the distance between the clock and the center of the DM object in the plane perpendicular to \v{\hat n}, the incident direction of the object.
In this plane, the distance of a clock from ECI0 is given by
\[{r^a_{\perp}}^2={\v{r}^a}^2 - (\v{\hat n}\cdot{\v{r}^a})^2.\]
If, in this plane, the DM object enters with a perpendicular distance of $R$ from ECI0, and at an angle $\alpha$ measured from  $z'$, the projection of $z$ down to the plane perpendicular to $\hat n$
$[\v{ \hat z}' = \v{\hat z} - (\v{\hat z}\cdot \v{\hat n})\v{\hat n}],$\footnotemark[4] 
 then the impact parameter is
\begin{equation}
{\rho^a} = \sqrt{{r^a_{\perp}}^2+R^2-2r^a_{\perp}R\cos{\gamma^a}},
\end{equation}
where ${\gamma^a}=\alpha-{\beta^a}$ is the angle between $R$ and $r^a_{\perp}$, and ${\beta^a}$ is the angle that $r^a_{\perp}$ makes in the plane perpendicular to $\v{\hat n}$ also measured from the $z'$-axis, and is given by
\[\tan{\beta^a}=\frac{({\v{r}^a}\times\v{\hat z})\cdot\v{\hat n}}{{\v{r}^a}\cdot\v{\hat z}-(\v{r}^a\cdot\v{\hat n})(\v{\hat n}\cdot\v{\hat z})},\]
 as shown in Fig.~\ref{fig:monopolegeom}.
The profile can be expressed as
\begin{equation}
\varphi^2(t^a,t')=\frac{v}{d \sqrt{\pi}}\exp\left( \frac{-v^2}{d^2}(t^a-t')^2 -\frac{{\rho^a}^2}{d^2} \right),
\label{eq:profile-monopole}
\end{equation}
and the parameter $h$ is linked to the field parameters as
\begin{equation}
\label{eq:h-monopole}
h = \frac{A^2 d \sqrt{\pi}}{v}
\sum_X{\k_X}\,{\Gamma_X}.
\end{equation}

\paragraph*{Strings---}

The string case is similar to the monopole case, except here the impact parameter is set by the perpendicular distance from each clock to the string.
We assume that on the scale of the GPS network, the string can be modelled as a straight line segment.
For a string that enters from incident direction  $\v{\hat n}_\perp=-\v{v_\perp}/v_\perp$ (we are interested in the velocity perpendicular to the string), with a perpendicular distance of $R$ from ECI0, at an angle $\alpha$ (measured from the $z'$-axis to $R$ as above), the impact parameter for each satellite is
\begin{equation}
{\rho^a} =R-r^a_{\perp}\cos{\gamma^a},
\end{equation}
where, as above, ${\gamma^a}=\alpha-{\beta^a}$ is the angle between $R$ and $r^a_{\perp}$, and ${\beta^a}$ is the angle that $r^a_{\perp}$ makes in the plane perpendicular to $\v{\hat n}_\perp$ also measured from the $z'$-axis.

Assuming a 2D Gaussian profile with radial width $d$, the string profile can be expressed
\begin{equation}
\varphi^2(t^a,t')=\frac{v_\perp}{d \sqrt{\pi}}\exp\left( \frac{-v_\perp^2}{d^2}(t^a-t')^2 -\frac{{\rho^a}^2}{d^2} \right).
\label{eq:profile-string}
\end{equation}
The parameter $h$ is linked back to the field parameters in the same way as for the Gaussian-profile wall case (\ref{eq:h-dwall}).

\footnotetext[4]{When $\abs{\v{\hat z}\cdot \v{\hat n}}\approx 1$, we instead define $\v{ \hat z}' = \v{\hat x} - (\v{\hat x}\cdot \v{\hat n})\v{\hat n}$.}

\bibliography{GPSDM,extra}

\end{document}